%% file: main.tex
\documentclass[a4paper,UKenglish,cleveref,autoref,thm-restate]{lipics-v2021}

\title{Polynomial prenexing of QBFs with non-monotone boolean operators} % equivalences}
\titlerunning{Polynomial prenexing of QBFs} % equivalences}
\author{Abdallah Saffidine}
{Sydney, Australia %\and \url{} 
}{abdallah.saffidine@gmail.com}
{https://orcid.org/0000-0001-9805-8291}
{}

\author{Andreas Herzig}
{IRIT, CNRS, France \and \url{https://www.irit.fr/~Andreas.Herzig} }{Andreas.Herzig@irit.fr}
{https://orcid.org/0000-0003-0833-2782}
{} %(Optional) author-specific funding acknowledgements}
\authorrunning{A. Saffidine and A. Herzig} 
\Copyright{Abdallah Saffidine and Andreas Herzig} %TODO mandatory, please use full first names. LIPIcs license is "CC-BY";  http://creativecommons.org/licenses/by/3.0/

 %TODO mandatory; please add comma-separated list of keywords
\category{} %optional, e.g. invited paper
\relatedversion{} %optional, e.g. full version hosted on arXiv, HAL, or other respository/website
\acknowledgements{We acknowledge discussions with several colleagues about the problem of prenexing with equivalences; in particular, an email exchange with Mikoláš Janota encouraged us to pursue our attempt. 
}

\EventEditors{John Q. Open and Joan R. Access}
\EventNoEds{2}
\EventLongTitle{42nd Conference on Very Important Topics (CVIT 2016)}
\EventShortTitle{CVIT 2016}
\EventAcronym{CVIT}
\EventYear{2016}
\EventDate{December 24--27, 2016}
\EventLocation{Little Whinging, United Kingdom}
\EventLogo{}
\SeriesVolume{42}
\ArticleNo{23}

\usepackage{cite}
\usepackage{amsmath,amssymb,amsfonts}
\usepackage{algorithmic}
\usepackage{graphicx}
\usepackage{textcomp}
\usepackage{xcolor}
\def\BibTeX{{\rm B\kern-.05em{\sc i\kern-.025em b}\kern-.08em
    T\kern-.1667em\lower.7ex\hbox{E}\kern-.125emX}}

\input{preamble}
\begin{document}

\title{Polynomial Prenexing of QBFs \\ with Non-Monotone Boolean Operators
%{\footnotesize \textsuperscript{*}Note: Sub-titles are not captured in Xplore and should not be used}
%\thanks{Identify applicable funding agency here. If none, delete this.}
}
\maketitle

\begin{abstract}
It is well-known that every quantified boolean formula (QBF) can be transformed into a prenex QBF whose only boolean operators are negation, conjunction, and disjunction. 
It is also well-known that the transformation is polynomial if the boolean operators of the original QBF are restricted to negation, conjunction, and disjunction. 
In contrast, up to now no polynomial transformation has been found when the original QBF contains other boolean operators such as biconditionals or exclusive disjunction. 
We define such a transformation and show that it
%We show how to transform QBFs with arbitrarily nested quantifiers, biconditionals, and other boolean operators into prenex form. The transformation 
is polynomial and preserves quantifier depth.  
\end{abstract}

%Keywords: Quantified boolean formulas, QBF, prenexing, complexity, polynomial hierarchy

% \author{anonymous} {anonymous affiliation} {} {} {}
% \authorrunning{anonymous} 
% \Copyright{anonymous}

% \author{Abdallah Saffidine}
% {Sydney, Australia %\and \url{} 
% }{abdallah.saffidine@gmail.com}
% {https://orcid.org/0000-0001-9805-8291}
% {}

% \author{Andreas Herzig}
% {IRIT, CNRS, France \and \url{https://www.irit.fr/~Andreas.Herzig} }{Andreas.Herzig@irit.fr}
% {https://orcid.org/0000-0003-0833-2782}
% {} %(Optional) author-specific funding acknowledgements}
% %TODO mandatory, please use full name; only 1 author per \author macro; first two parameters are mandatory, other parameters can be empty. Please provide at least the name of the affiliation and the country. The full address is optional. Use additional curly braces to indicate the correct name splitting when the last name consists of multiple name parts.
% \authorrunning{A. Saffidine and A. Herzig} 
% \Copyright{Abdallah Saffidine and Andreas Herzig} %TODO mandatory, please use full first names. LIPIcs license is "CC-BY";  http://creativecommons.org/licenses/by/3.0/

\allowdisplaybreaks %to allow page breaks inside math environments globally.

%%%%%%%%%%%%%%%%%%%%%%%%%%%%%%%%%
\section{Introduction}\label{sec:intro}
%%%%%%%%%%%%%%%%%%%%%%%%%%%%%%%%%

%\ah{'biconditional' au lieu de 'equivalence' pour eviter toute confusion possible avec 'logical equivalence' (cf. l'email de Janota ou il dit "...you mean 'biimplication'?" }
Provers for Quantified Boolean Formulas (QBF) require input formulas to be in prenex form: 
sequences of quantifiers over propositional variables (the prenex) followed by a boolean formula in conjunctive normal form CNF (the matrix). 
Any QBF whose  boolean operators are restricted to negation, disjunction, and conjunction can be put in that form:
one can move quantifiers outwards across these operators, replacing e.g.\ subformulas
%$\chi \lor \exists x \phi$ by $\exists x (\chi \lor \phi) $, 
$\chi \land \exists x \phi$ by $\exists x (\chi \land \phi) $.
% \begin{align*}
% \lnot \exists x \phi &\leqv \forall x \lnot \phi , 
%   &\lnot \forall x \phi &\leqv \exists x \lnot \phi , 
% \\
% \phi \lor \exists x \psi &\leqv \exists x (\phi \lor \psi) ,
%   &\phi \lor \forall x \psi &\leqv \forall x (\phi \lor \psi) ,
% \\
% \phi \land \exists x \psi &\leqv \exists x (\phi \land \psi) ,
%   &\phi \land \forall x \psi &\leqv \forall x (\phi \land \psi) .
% \end{align*}
The resulting formula is logically equivalent to the original formula if $x$ does not occur free in $\chi$, which can always be ensured by renaming. %before moving the quantifier outwards. 
By iterating the application of these equivalences a formula in prenex form is obtained whose length is linear in the length of the original formula. 

Moving quantifiers outwards is less obvious when formulas contain other boolean operators. 
The reason is that when a formula only contains negation, disjunction, and conjunction then one can associate a unique polarity to each subformula; 
this fails to hold when the formula contains operators that are non-monotone, such as the biconditional $\leqv$ and the exclusive disjunction $\oplus$. % or the exclusive disjunction operator. 
One can still move quantifiers outwards by eliminating the operators in a preprocessing step. 
For example, $\chi\leqv \exists x \phi$ is first rewritten to
$(\chi \land \exists x \phi) \lor (\lnot\chi \land \forall x \lnot\phi)$; 
then $x$ is renamed to $x^-$ in the second conjunct $\forall x \lnot \phi$, resulting in
$(\chi \land \exists x \phi) \lor (\lnot\chi \land \forall \cpn x \lnot \phi [\{x/ \cpn x\}] )$; 
finally both quantifiers are prenexed. 
This, however, leads to exponential growth when non-monotone operators are nested. 

%Languages where equivalences are primitive are of interest e.g.\ when decision problems are translated into QBFs. 
%For example, the upper bound of the complexity of the logic with the epistemic concept of an agent only knowing $\phi$ \cite{DBLP:journals/ai/Levesque90} can be obtained via a transformation into the language of QBFs with equivalence \cite{DBLP:conf/jelia/LimaLS23}. 

One solution is to implement solvers accepting formulas that are not in prenex form; see e.g.\ \cite{DBLP:journals/tcad/GiunchigliaNT07,DBLP:journals/constraints/EglySW09,DBLP:conf/sat/KlieberSGC10,DBLP:conf/sat/Tentrup16,DBLP:conf/ijcar/HeisingerHRS24}.
%, or adapt existing solvers in order to keep track of dependences 
We here take another route and define a polynomial transformation into prenex form. 
In the past several such transformations were introduced and studied in the literature \cite{DBLP:conf/sat/EglySTWZ03,DBLP:conf/ecai/EglySW06,DBLP:conf/sat/LonsingB10,DBLP:conf/icaart/Stephan14}. 
They improve over the naive preprocessing-based algorithm we have sketched above. 
Nevertheless, when formulas contain nested quantifiers and biconditionals then the resulting prenex formulas are exponentially bigger in the worst case. 
To the best of our knowledge, no polynomial prenexing transformation is known when the language contains biconditionals. % that preserves quantifier depth. 

\goodbreak
In this paper we show how one can put QBFs containing biconditionals---and more generally $n$-ary boolean operators---%as well as quantifiers governed by boolan operators 
in prenex form. 
Our transformation draws inspiration from Tseytin's transformation for propositional logic~\cite{Tseytin}
and its generalisation to first-order predicate logic by Plaisted and Greenbaum~\cite{DBLP:journals/jsc/PlaistedG86}.
(More recently, another polynomial transformation has been proposed by Harwath \cite{DBLP:journals/ipl/Harwath16}.)
Contrarily to what one may expect, it is not obvious to adapt the Plaisted-Greenbaum transformation from predicate logic formulas to QBFs. 
Indeed, it makes use of fresh predicates in order to abbreviate subformulas, and it is essential that these predicates have object variables as arguments: there is no counterpart for such a device in the language of QBFs. 
\ah{reste ajoute}
Let us illustrate this by means of the formula 
$\exists z 
( (\exists x ((x \leqv z) \land \lnot x)) \leqv (\exists y ((y \leqv z) \land y) ) )$. 
A naive adaptation of the Plaisted-Greenbaum transformation would result in 
$\exists zxypq ( 
(p \leqv ((x \leqv z) \land \lnot x)) \land 
(q \leqv ((y \leqv z) \leqv y)) \land 
(p \leqv q) )$, 
with fresh variables $p$ and $q$.
However, the latter fails to be satisfiability-equivalent to the former. 

Let us illustrate our transformation by the formula
$$ \chi\oplus \exists x \phi$$ 
where $x$ does not occur in $\chi$. 
We first apply  Shannon-expansion,
that is, we rewrite the formula so that we distinguish the two possible truth values that $\exists x \phi$ can take:
$$(\exists x \phi \land (\chi\oplus\top)) \lor (\forall x \lnot \phi \land (\chi\oplus\bot)) . $$
We then rename one of the variables in the two quantifications over $x$ by introducing fresh variables $\cpp x$ and $\cpn x$,
%so that $\exists \cpp x$ and $\forall \cpn x$ can be prenexed, 
resulting in 
$$\left( \exists \cpp x \exists x \, (\phantom{\lnot} \phi \land (x\leqv \cpp x) \land (\chi\oplus\top)) \right) 
\lor\ 
\left( \forall \cpn x \exists x \, (\lnot \phi \land (x\leqv \cpn x) \land (\chi\oplus\bot)) \right) . 
$$
The latter can then be prenexed to
$$\forall \cpn x \exists \cpp x \exists x ((\phi \land (x\leqv \cpp x) \land (\chi\oplus\top)) \lor (\lnot \phi \land (x\leqv \cpn x) \land (\chi\oplus\bot)) ) , $$
which is propositionally equivalent to
$$\forall \cpn x \, \exists \cpp x \exists x \, \left(
\left(       \phi \limp (x\leqv \cpp x) \right) \land
\left( \lnot \phi \limp (x\leqv \cpn x) \right)
\land (\chi \oplus \phi) 
\right) . 
$$
Finally, we abbreviate $\phi$ by a fresh $p$, resulting in  
$$
\forall \cpn x \, \exists \cpp x \exists x \exists p \, \big(
(p \leqv \phi) 
\land \left(       p \limp (x\leqv \cpp x) \right) 
\land \left( \lnot p \limp (x\leqv \cpn x) \right) 
\land\ (\chi \oplus p)
\big) .
$$
%
% $$
% \forall \cpn x \, \exists x \exists \cpp x \exists p \, \left(
% (p \leqv \phi) \land 
% \left(       p \limp (x\leqv \cpp x) \right) \land
% \left( \lnot p \limp (x\leqv \cpn x) \right)
% \land (\chi \leqv p)
% \right) .
% $$
% Let us illustrate it by the above
% $ \chi\leqv \exists x \phi . $ 
% We first rewrite it to 
% $$(\exists x \phi \land (\chi\leqv\top)) \lor (\forall x \lnot \phi \land (\chi\leqv\bot)) . $$
% We then distinguish the two quantifications over $x$ by introducing fresh variables $\cpp x$ and $\cpn x$ so that $\exists \cpp x$ and $\forall \cpn x$ can be prenexed, resulting in 
% $$\forall \cpn x \exists x \exists \cpp x ((\phi \land (x\leqv \cpp x) \land (\chi\leqv\top)) \lor (\lnot \phi \land (x\leqv \cpn x) \land (\chi\leqv\bot)) ) , $$
% which is propositionally equivalent to
% $$\forall \cpn x \, \exists x \exists \cpp x \, \left(
% \left(       \phi \limp (x\leqv \cpp x) \right) \land
% \left( \lnot \phi \limp (x\leqv \cpn x) \right)
% \land (\chi \leqv \phi) 
% \right) . 
% $$
% Finally, we abbreviate $\phi$ by a fresh $p$, resulting in  
% $$
% \forall \cpn x \, \exists x \exists \cpp x \exists p \, \left(
% (p \leqv \phi) \land 
% \left(       p \limp (x\leqv \cpp x) \right) \land
% \left( \lnot p \limp (x\leqv \cpn x) \right)
% \land (\chi \leqv p)
% \right) .
% $$
The last formula is logically equivalent to the initial formula $\chi \oplus \exists x \phi$ under the proviso that $x$ does not occur free in $\chi$. 

In this paper we generalise the above transformation to quantified boolean formulas where quantifiers and $n$-ary boolean operators are arbitrarily nested. 
We call the language of such formulas \emph{full QBF}. 
We define full QBFs and their semantics in Section~\ref{sec:qbf}. 
We then introduce our transformation: 
we first prenex outermost quantifiers (Section~\ref{sec:trafo_level_zero}) and then all of them (Section~\ref{sec:trafo_any_level}). 
%; we prove that the transformation is correct, preserves quantifier depth, and only comes with a linear increase in formula length. 
In Section~\ref{sec:discu} we discuss implementation and evaluation of our transformation and Section~\ref{sec:conclu} concludes.

%%%%%%%%%%%%%%%%%%%%%%%%%%%%%%%%%
\section{Full QBFs}\label{sec:qbf}
%%%%%%%%%%%%%%%%%%%%%%%%%%%%%%%%%

Our language of full QBFs has \emph{quantifier blocks}: quantifiers applying to  finite sets of propositional variables. 
These quantifiers can occur in the scope of boolean operators. The latter are arbitrary: 
we suppose given some set of boolean functions $\boolfunsemantics k$, each of some arity $k \geq 0$, 
whose syntactical counterpart are boolean operators noted $\boolfunname{k}$.
We suppose that the set of these operators contains 
%the zero-ary operator $\bot$ and the binary $\limp$
the zero-ary operator $\bot$, the unary $\lnot$, and the binary $\land$, $\limp$, and $\leqv$. 
%$\top$ %the binary $\limp$, $\lor$ , 
%\ah{a verifier: eske "a complete set of boolean operators containing $\leqv$ " marche aussi ?}

\goodbreak 

%-----------------------------------------------------------------%
\subsection{Language}

Given a countably infinite set of propositional variables $\propset$, the language $\lang{}$ of full QBFs is formally defined to be the smallest set such that:
\begin{itemize}
\item 
$p \in \lang{}$ for every $p \in \propset$; 
\item 
$\boolfun k {\phi_1,\ldots,\phi_k} \in \lang{}$ if $\phi_1,\ldots,\phi_k \in \lang{}$ and $\boolfunname k$ is a $k$-ary boolean operator; 
\item 
$\exists X \phi \in \lang{}$ if $\phi \in \lang{}$ and $X$ is a finite subset of $\propset$; 
\item 
$\forall X \phi \in \lang{}$ if $\phi \in \lang{}$ and $X$ is a finite subset of $\propset$. 
\end{itemize}
We say that $\exists$ is the dual of $\forall$, and vice versa. 
Note that in $\forall X \phi$ and $\exists X \phi$ the set $X$ can be empty. 
We write $\exists x \phi$ instead of $\exists \{x\} \phi$ and $\forall x \phi$ instead of $\forall \{x\} \phi$. 
%For the sake of notational economy our language has no universal quantifiers: we are going to use $\forall X \phi$ as an abbreviation of $\lnot\exists X \lnot \phi$. 
%We suppose that $\phi \limp \psi$ is an abbreviation of $\lnot(\phi \land \lnot \psi)$. 
%
The abbreviation $\bigwedge_{k \leq n} \phi_k$ stands for the conjunction $\phi_1 \land \cdots \land \phi_n$, where we suppose that it equals $\top$ when $n$ is zero. 

The set of boolean formulas $\lang{bool}$ is the set of full QBFs from $\lang{}$ without $\exists$ and $\forall$. 
We use $\beta, \beta',\ldots$ to denote boolean formulas. 
% by the following grammar:
% \[\begin{array}{lcl}
% \lang{} ~:~
% \phi  \bnf p \mid f^n(\phi,\ldots,\phi)
%                         \mid \exists X \phi \mid \forall X \phi 
% \end{array}\]
% where $p$ ranges over a countably infinite set of propositional variables $\propset$, 
% $n\geq 0$ ranges over the integers, and
% $X$ ranges over the set of finite subsets of $\propset$. 
%We drop set parentheses and simplify quantifier blocks such as $\exists \{ p_1,\ldots,p_n\}$ to $\exists p_1 \ldots p_n$.
%In the QBF language $\lang{RQBF}$ it will be convenient to generalise quantification over variables to quantification over sets of variables: for $X = \{p_1,\ldots,p_n\} \subseteq \propset$, the formula $\exists X \phi $ abbreviates $\exists p_1 \cdots \exists p_n \phi$ (the order does not matter). 
%the boolean operators are negation, disjunction, conjunction, and equivalence is a special case. 

%The set of \emph{subformulas} of a formula $\phi$ is noted $\subfml\phi$.

%-----------------------------------------------------------------%
\subsection{Bound Variables, Free Variables, and Substitutions}

%Given a full QBF $\phi \in \lang{}$, 
The \emph{set of propositional variables occurring in} $\phi$ is noted $\propsetOf \phi$; % = \propset\cap \subfml\phi$. 
its set of \emph{free variables} is $\freevars{\phi}$ and its set of \emph{bounded variables} is $\boundvars{\phi}$. 
The \emph{number of blocks} in $\phi$, noted $\nblock\phi$, is the number of occurrences of quantified subformulas $Q X \psi$ in $\phi$. 
%= \card{ \{\psi \in \subfml\phi \suchthat \psi \text{ is of the form } Q X \psi'\} } $.  
The \emph{number of quantified variables} in $\phi$, noted $\nbvar\phi$, is the sum of the lengths of the occurrences of blocks in $\phi$. 
%= \sum_{Q_i X_i \in \subfml\phi} \card{X_i}$.
For example, for the formula 
$\phi = \exists x (\psi \land \lnot \exists \{x,y\} \chi) \land \lnot \forall y \rho$ 
we have $\nblock\phi = 3$ and $\nbvar\phi = 4$.

A \emph{substitution} is a set of couples 
$\sigma = \{p_1/\phi_1,\ldots,p_k/\phi_k\}$ 
such that $p_i \neq p_j$ for $i \neq j$.  
Its application to a formula $\phi$, noted $\phi[\sigma]$, simultaneously replaces all free occurrences of the variables $p_i$ in $\phi$ by $\phi_i$. 
Instead of 
$\phi [\{p_1/\phi_1,\ldots,p_k/\phi_k\}]$ we sometimes write 
$\phi [p_1/\phi_1,\ldots,p_k/\phi_k]$ or
$\phi [p_i/\phi_i \suchthat i\leq k]$. 
We suppose that substitutions have narrow scope. 
For example, $\phi \land \psi[\sigma]$ stands for $\phi \land (\psi[\sigma])$.
% is defined by:
% \begin{align*}
% p[x_1/\phi_1,\ldots,x_k/\phi_k] &= \begin{cases} \phi_i &\text{if } p=x_i \text{ for some } i\\ 
%                                                   p &\text{otherwise} % \end{cases} \\
% \exists X \phi[x_1/\phi_1,\ldots,x_k/\phi_k] &= \begin{cases} % \exists X \phi &\text{if } x_i \in X \text{ for some } i\\ 
%                                                   \exists X (\phi[x_1/\phi_1,\ldots,x_k/\phi_k]) &\text{otherwise} \end{cases}
% \end{align*}
% and homomorphic otherwise. 

%-----------------------------------------------------------------%
\subsection{Metrics}

The \emph{length} of a formula $\phi \in \lang{}$, noted $\lngth \phi$, is the number of logical and non-logical symbols occurring in $\phi$, where 
propositional variables and logical operators both have length $1$ and negations and parentheses do not count. 
Formally, the inductive definition is:
$\lngth{p} = 1$; 
$\lngth{\mathbf f^n(\phi_1,\ldots,\phi_n)} =  1+ \sum_{i \leq k} \lngth{\phi_i}$; and
%$\lngth{\mathbf f^n(\phi_1,\ldots,\phi_n)} =  1+ \lngth{\phi_1} + \ldots + \lngth{\phi_n}$; and
$\lngth{\exists X \phi} = %1 + \card X + \lngth{\phi}$; and $
 \lngth{\forall X \phi} = 1 + \card X + \lngth{\phi}$. 
For example, the length of $p \leqv q$ is $3$ and the length of $\exists \{x,y\} \ (x \limp p)$ is 
$1 + \card{\{x,y\}} + \lngth{ x \limp p } = 1 + 2 + 3 = 6$. 
%\lngth{ (\lnot p \lor q) \land (p \lor \lnot q)} = 1 + 4 + 4 = 9 .

%-----------------------------------------------------------------%
%\subsection{Quantifier Depth} 
\emph{Quantifier depth} of $\phi$, noted $\mdepth{\phi}$, is defined by: 
\begin{align*}
\mdepth{p} &= 0 ; 
\\
\mdepth{\mathbf f^n(\phi_1,\ldots,\phi_n)} &=  \max \big(\mdepth{\phi_1},\ldots,\mdepth{\phi_n} \big) ; 
\\
\mdepth{\exists X \phi} &= % 1+\mdepth{\phi}$, and
 \mdepth{\forall X \phi} = 1+\mdepth{\phi} ; 
\end{align*}
where we suppose that $\max \big(\mdepth{\phi_1},\ldots,\mdepth{\phi_n} \big)$ equals $0$ when $n$ equals $0$. %$\mathbf f^n$ is a $0$-ary boolean operator such as $\bot$. 
Hence the quantifier depth of a formula is zero if and only if it is boolean. 
%The quantifier depth of a formula in prenex form $\Lambda \beta$  is the number of occurrences of $\exists$ in $\Lambda$. 

Our definition of quantifier depth differs from the more standard notion of alternation depth: 
there are formulas with $\mdepth \phi > 1$ where $\exists$ and $\forall$ do not alternate, such as $\phi=\exists X \exists Y \psi$. 
The reason for our choice is that quantifier depth is a more natural notion when there are non-monotone boolean operators. 
Indeed, when there are only monotone boolean operators then it makes sense to say that the depth of e.g.\ 
$\exists x \lnot \exists y x \lor y$ should be 2 because to each subformula one can associate a unique polarity. 
However, it is more difficult to associate a meaningful notion of an alternation depth to formulas such as 
$\exists x ((\exists y y) \leqv x)$. 
%\abda{Je dirais plutôt qu'on utilise quantifier deepth parce que c'est une notion plus naturelle en présence d'opérateurs non-monotones: e.g., ça va à peu près de dire que $\exists x \lnot \exists y x \lor y$ est d'alternation depth 2 en utilisant la notion de polarité des sous-formules, mais quand on regarde $\exists x ((\exists y y) \leqv x)$, l'intuition d'alternation depth commence à devenir trop maladroite.}

%-----------------------------------------------------------------%
\subsection{Prenex Normal Form}

A \emph{prefix} is a sequence of $\exists X$ and $\forall Y$. 
%of quantifiers $Q_1 X_1 \cdots Q_n X_n$ where each $Q_i$ is either $\exists$ or $\forall$.
%We use $\Lambda$, $\Lambda'$, etc.\ to denote prefixes; the empty prefix (the empty sequence) is noted $\emptyprefix$. 
A QBF is in \emph{prenex form} if it is of the form $\Lambda \beta$ where $\Lambda$ is a prefix and $\beta \in \lang{bool}$. % has no quantifier. 
Hence the quantifier depth of a formula in prenex form is the number of occurrences of $\exists$ and $\forall$ in the prefix.
%We note that more formal exposition would require a grammar where sequences of quantifier blocks are first-class citizens. We have chosen to avoid this for the sake of readability.

%-----------------------------------------------------------------%
\subsection{Semantics}\label{sec:qbf_semantics}

A \emph{valuation} is a mapping from $\propset$ to $\{0,1\}$.
%When $p \in \val$ then $p$ is true in $\val$ and 
%when $p \not\in \val$ then $p$ is false in $\val$.
%hence the set of all valuations is $2^\propset$. 
We use $\val,\valb,\ldots$ for valuations. 

We say that two valuations $\val$ and $\val'$ \emph{agree} on a set of propositional variables $X \subseteq \propset$ if their restrictions to $X$ are identical, that is, if $\val|_X = \val'|_X$. 
%$\val \cap X = \val' \cap X$. 

Valuations are extended from $\propset$ to complex formulas of $\lang{}$ as follows:
%\abda{L'évaluation d'une proposition a l'air d'une typo / d'un formalisme différent.}
%$\intFml{\bot} = \emptyset$; 					%\intFml{\top} =  2^{\propset}
%$\intFml{\lnot \phi} = 2^\propset \setminus \intFml{\phi}$; 
%$\intFml{\phi \lor \psi} = \intFml{\phi} \cup \intFml{\psi}$; 
\begin{align*}
%\intFml{p} &= \{\val \suchthat p \in \val \} ; \\
\intFml{\boolfun k {\phi_1,\ldots,\phi_k} } &= 
\boolfunsemantics k(\intFml{\phi_1},\ldots,\intFml{\phi_k}) ;
%\\
%\intFml{\mathbf{f}(\phi_1,\ldots,\phi_k) } &= 
%\{ V \suchthat f(b_1,\ldots,b_k) \text{ where } b_i = \top \text{ iff } V \in \intFml{\phi_i}\};
\\
\intFml{\exists X \phi} &= \begin{cases} 
1 &\text{if } V'(\phi) = 1 \text{ for some } V' \text{ agreeing with } V \text{ on } \propset \setminus X, \\
0 &\text{otherwise} ;
\end{cases}
\\
\intFml{\forall X \phi} &= \begin{cases} 
1 &\text{if } V'(\phi) = 1 \text{ for every } V'  \text{ agreeing with } V \text{ on } \propset \setminus X, \\
0 &\text{otherwise} .
\end{cases}
\end{align*}

A formula $\phi \in \lang{}$ is \emph{satisfiable} if $\intFml \phi = 1$ for some valuation $\val$; 
it is \emph{valid} if $\intFml \phi = 1$ for every $\val$. 
Two formulas $\phi$ and $\psi$ are \emph{logically equivalent}, written $\phi \equiv \psi$, if $\phi\leqv \psi$ is valid; that is, if $\intFml \phi = \intFml \psi$ for every valuation $\val$. 
They are \emph{equisatisfiable} if $\phi$ is satisfiable exactly when $\psi$ is satisfiable; and 
they are \emph{equivalid} if $\phi$ is valid exactly when $\psi$ is valid. 

%-----------------------------------------------------------------%
%\subsection{Some Useful Properties}

We state some logical equivalences that are going to be useful. 
Remember that a formula $\psi$ is \emph{free (to be substituted) for} $p$ in $\phi$ if it is not the case that there is a free occurrence of $p$ in $\phi$ that is in the scope of a quantifier in $\phi$ binding some free variable of $\psi$. 
%For example, $x \land q$ is not free for $p$ in $\exists x p$. 
A particular case is when no occurrence of $p$ in $\phi$ is in the scope of a quantifier: 
then any $\psi$ is free for $p$ in $\phi$. 

\medskip\begin{fact}\label{lemma:subst} 
Let $\phi$ be a full QBF in which $\psi$ is free for $p$. Then: % is not in the scope of any quantifier. 
%let $\{p / \psi \}$ be a substitution. 
\begin{enumerate}
%\item\label{lemma:subst_freefor}  
%If $\phi \equiv \phi'$ and $\psi$ is free for $p$ in $\phi'$ then $\phi[p / \psi] \equiv \phi'[p / \psi]$; 
% \item\label{lemma:subst_truefalse} 
% $\phi \equiv (\phi[p/\top] \land p) \lor (\phi[p/\bot] \land \lnot p)$. 
\item\label{lemma:subst_cases}  
$\phi[p/\psi] \equiv (\phi[p/\top]\land \psi) \lor (\phi[p/\bot] \land \lnot \psi) $; 
\item\label{lemma:subst_equiv} 
If $p \notin \freevars\psi$ % does not occur in $\psi$
% \ah{manquait avant ; contre-exemple sinon : 
% $\phi = \psi = p$
% %($p[p/p] = p$ n'est pas equivalent a $\exists p((p \leqv p) \land p))$ }
then $\phi[p/\psi] \equiv \exists p \left( (p \leqv\psi) \land \phi \right) $. 
\end{enumerate}
\end{fact} 
\begin{pf}
%Item~\ref{lemma:subst_freefor} is standard. 
For item~\ref{lemma:subst_cases}, 
$\phi$ is logically equivalent to 
$(\phi[p/\top] \land p) \lor (\phi[p/\bot] \land \lnot p)$. 
As $\psi$ is free for $p$ in both formulas, 
%(because $p$ is not in the scope of any quantifier in $\phi$),
the substitution $\{p / \psi\}$ preserves this: % by Item~\ref{lemma:subst_freefor}: 
$\phi[p/\psi]$ is logically equivalent to 
$( (\phi[p/\top] \land p) \lor (\phi[p/\bot] \land \lnot p) ) [p/\psi]$. 
The latter is logically equivalent to 
$(\phi[p/\top]\land \psi) \lor (\phi[p/\bot] \land \lnot \psi) $.

For item~\ref{lemma:subst_equiv}, according to the truth condition for $\exists$ the right hand side 
$\exists p \left( (p \leqv\psi) \land \phi \right) $ 
is logically equivalent to the disjunction
$\left( \phi \land (p \leqv\psi) \right)[p/\top] \lor 
\left( \phi \land (p \leqv\psi) \right) [p/\bot] $.
The latter equals
$\left( \phi[p/\top] \land (\top \leqv \psi) \right) \lor  
\left( \phi[p/\bot] \land (\bot \leqv \psi) \right) $
because $p$ does not occur free in $\psi$; 
which simplifies to 
$\left( \phi[p/\top] \land \psi \right) \lor  
\left( \phi[p/\bot] \land \lnot\psi \right) $. 
This is logically equivalent to the left hand side $\phi[p/\psi]$ by item~\ref{lemma:subst_cases}. 
%(which applies because of the hypothesis that $\psi$ is free for $p$ in $\phi$).
%The condition that $p$ is not in the scope of any quantifier in $\phi$ is necessary: otherwise e.g.\ the formulas $(\exists x p) [p/x] = \exists x x$ and $  p [p/x] = x$ would be equivalent, which is false. 
\end{pf}

\begin{fact}\label{lemma:compo_subst}
Let $\sigma = \{p_1/\phi_1, \ldots, p_k/\phi_k\}$
be a substitution. 
Let $p \in \propset$ such that 
%If $p$ does not occur in $\sigma$\footnote{More precisely: 
for every $ i \leq k$ , %$p_i/\phi_i \in \sigma$, 
$p \notin \{p_i\} \cup \freevars{\phi_i}$. 
Then 
$\phi[\sigma \cup \{p/\psi\}] \equiv 
\phi[\sigma] [p/\psi]$.
%$\phi[p_1/\phi_1, \ldots, p_k/\phi_k , p_{k+1}/\phi_{k+1}] \equiv (\phi[p_1/\phi_1, \ldots, p_k/\phi_k]) [p_{k+1}/\phi_{k+1}]$.
\end{fact}

\medskip\begin{fact}\label{lemma:freshvariables_equiv}
Let $\forall X \phi$ be a full QBF. 
Let $\cpn X = \{\cpn x \suchthat x \in X\}$ %and $\cpp X = \{\cpp x \suchthat x \in X\}$
be fresh. Then
$$\forall X \phi \equiv \forall \cpn X \exists X \left( \phi \land \bigwedge_{x \in X}(x {\leqv} \cpn x) \right) . $$
\end{fact}

% \begin{fact}[univ2]\label{lemma:freshvariables_equiv}
% Let $\forall X \phi$ be a full QBF. 
% Let $\cpn X = \{\cpn x \suchthat x \in X\}$ and $\cpp X = \{\cpp x \suchthat x \in X\}$ be fresh. Let $\sigma = \{x/\cpp x \suchthat x \in X\}$ Then: 
% \begin{align*}
% \forall X \phi &\equiv \forall \cpn X \exists \cpp X \left( \phi[\sigma] \land\bigwedge_{x \in X}(x \leqv \cpn x) \right) ;
% \\
% \exists X \phi &\equiv \forall \cpn X \exists \cpp X \left( \phi[\sigma] \land \bigwedge_{x \in X}(x \leqv \cpp x) \right) .
% \end{align*}
% \end{fact}

%-----------------------------------------------------------------%
\subsection{Decision Problems}

The satisfiability problem is to decide whether a given formula $\phi \in \lang{}$ is satisfiable; 
%whether $\intFml \phi = 1$ for some $\val$; 
and the validity problem is to decide whether $\phi$ is valid. 
The model checking problem is to decide, given a valuation $\val$ and a formula $\phi \in \lang{}$, whether $\intFml \phi = 1$. 

It is known that for QBFs in prenex form, the satisfiability problem for formulas of quantifier depth $k$ is $\Sigma_{k+1}^{\ptime}$-complete and the validity problem is $\Pi_{k+1}^{\ptime}$-complete. 
Our transformation will establish that the same is the case for full QBFs.
%\abda{J'ai écrit la footnote ci-dessous en réponseà ta question avant d'avoir lu la conclusion. Du coup, si tu veux, on peut ne pas parler de collapse ici et mettre la référence à Hemaspaandra et al dans la conclusion (et en profiter pour être plus précis "collapses to $k$-th level).}
It will moreover establish that the model checking problem for full QBFs of quantifier depth $k$ is $\Theta_k^P$-complete. 
This contrasts with prenex QBFs, for which model checking is not $\Theta_k^P$-hard under standard computational complexity assumptions.

%%%%%%%%%%%%%%%%%%%%%%%%%%%%%%%%%
\section{Prenexing Outermost Quantifiers}\label{sec:trafo_level_zero}
%\section{From full QBF to Prenex QBF}\label{sec:trafo}
%%%%%%%%%%%%%%%%%%%%%%%%%%%%%%%%%

An outermost quantifier is not in the scope of any other quantifier. 
The first step of our transformation is to prenex all such quantifiers without doubling the occurrences of the quantified subformula. 
This requires some fresh auxiliary variables. 

We start by a lemma that moves one quantifier outwards and fuses it with an existing quantifier $\exists Y$ (for which $Y$ may be empty). 

\medskip\begin{lemma}\label{lemma:extract_quantifier} 
Let $\psi, \phi \in \lang{}$ be full QBFs. % \in \lang{}$. 
Let $p \in \propsetOf{\psi} \setminus \freevars{\phi}$ be not in the scope of any quantifier in $\psi$. 
% \ah{avant: "$p \in \propsetOf{\psi}$", mais comme R3 a remarqué, cans la derniere equivalence on utilise que $p \notin Free (\varphi[\sigma])$:\\
% "proof of Lemma 1:\\
% - $p \notin Free (\varphi[\sigma])$ is not guaranteed, since $p$ could be in $Free(\varphi) \setminus X$\\
% - $p$ is basically a name given to the position of $QX\varphi$ in the formula\\
% - so demanding $p \notin \varphi$ should be possible"\\
% Il faut rajouter ca comme hypothese ; comme le dit R3 on peut toujours la faire. 
% }
%$p$ does not occur in $\psi$ and 
%$X \subseteq \propsetOf{\psi} \setminus \propsetOf{\phi}$. 
Let $Y\subseteq \propsetOf{\psi} \setminus \propsetOf{\phi}$. 
%and $\cpp X = \{\cpp x \suchthat x \in X\}$ 
%Let $\cpp \sigma = \{ x / \cpp x \suchthat x \in X \}$.
Let $\cpn X = \{\cpn x \suchthat x \in X\}$ and $\cpp X = \{\cpp x \suchthat x \in X\}$ be sets of fresh variables. 
Let $\sigma = \{x/\cpp x \suchthat x \in X \}$.
%Let $Q$ be either $\exists$ or $\forall$.
Then 
\begin{align*}
\exists Y \left(\psi[p/\exists X \phi]\right) &\equiv 
\forall \cpn X\;\exists Y {\cup} \cpp X {\cup} \{p\} \left( 
\begin{aligned}
\psi \land (p \leqv \phi[\sigma]) %\\ 
\land 
\left( \bigwedge_{x \in X}(\lnot p \limp (\cpp x {\leqv} \cpn x)) \right) 
\end{aligned}
\right) ;
\\
\exists Y \left(\psi[p/\forall X \phi] \right)&\equiv 
\forall \cpn X\;\exists Y {\cup} \cpp X {\cup} \{p\} \left( 
\begin{aligned}
\psi \land (p \leqv \phi[\sigma]) %\\ 
\land
\left( \bigwedge_{x \in X}(\phantom{\lnot} p \limp (\cpp x {\leqv} \cpn x)) \right) 
\end{aligned}
\right) .
\end{align*}
% \begin{align*}
% \exists Y \left(\psi[p/Q X \phi]\right) &\equiv 
% \begin{cases}
% \forall \cpn X\;\exists Y {\cup} \cpp X {\cup} \{p\} \; \left( 
% (p \leqv \phi[\sigma]) \land 
% \left( \bigwedge_{x \in X}(\lnot p \limp (\cpp x {\leqv} \cpn x)) \right) \land \psi
% \right) & \text{if $Q = \exists$,}
% \\
% %\exists Y \left(\psi[p/Q X \phi] \right)&\equiv 
% \forall \cpn X\;\exists Y {\cup} \cpp X {\cup} \{p\} \; \left( 
% (p \leqv \phi[\sigma]) \land 
% \left( \bigwedge_{x \in X}(\phantom{\lnot} p \limp (\cpp x {\leqv} \cpn x)) \right) \land \psi
% \right) & \text{if $Q = \forall$.}
% \end{cases}
% \end{align*}
\end{lemma}

\begin{pf}
For the first equivalence we give a sequence of formulas whose first is the left side and whose last is the right side and which are all logically equivalent: 
\begin{enumerate}
\item $ %1
\exists Y \, \left(  \psi[p/\exists X \phi] \right)
\smallskip $ \item $ %2
\exists Y \, \left( (\psi[p/\top] \land \exists X \phi) \lor 
(\psi[p/\bot] \land \lnot \exists X \, \phi ) \right)
$\\ \phantom x \hfill 
by Fact~\ref{lemma:subst}.\ref{lemma:subst_cases} (applies: $p$ is not in the scope of any quantifier in $\psi$, so $\exists X \phi$ is free for $p$ in $\psi$)
\smallskip \item $ %3
\left( \exists Y \, (\psi[p/\top] \land \exists X \phi) \right)
\lor
\exists Y \, \left( \psi[p/\bot] \land \forall X \lnot \phi \right) 
$ \hfill distribution of $\exists Y$ over $\lor$
\smallskip \item $ %4
\left( \exists Y (\psi[p/\top] \land \exists X \phi) \right) 
\lor
\left( (\exists Y \psi[p/\bot]) \land \forall X \lnot \phi \right) 
$ \hfill distribution of $\exists Y$ over $\land$ \\ \phantom x \hfill (correct because $Y \cap \freevars{\phi} = \emptyset$)
\smallskip \item $ %5
\left( \exists Y (\psi[p/\top] \land \exists X \phi) \right) 
\lor %$\\$
\left( (\exists Y \psi[p/\bot]) \land \forall \cpn X \, \exists X \left(\lnot \phi \land \bigwedge_{x \in X}(x {\leqv} \cpn x) \right) \right) 
$ \hfill by Fact~\ref{lemma:freshvariables_equiv}, for fresh $\cpn X$
\smallskip \item $ %6
\left( \exists Y (\psi[p/\top] \land \exists \cpp X \phi[\sigma]) \right) 
\lor 
\left( (\exists Y \psi[p/\bot]) \land \forall \cpn X \, \exists \cpp X \left(\lnot \phi[\sigma] \land \bigwedge_{x \in X}(\cpp x {\leqv} \cpn x) \right) \right) 
$ \\ \phantom x \hfill by definition of $\sigma$
\smallskip \item $ %7
\forall \cpn X \; \left( \begin{aligned}
(\exists Y (\psi[p/\top] \land \exists \cpp X \phi[\sigma] )) %\\
\lor
\left( (\exists Y \psi[p/\bot]) \land 
\exists \cpp X \left( \lnot \phi[\sigma] \land \bigwedge_{x \in X}(\cpp x {\leqv} \cpn x) \right) \right) 
\end{aligned} \right) 
$\\ \phantom x \hfill distribution of $\forall \cpn X$ over $\land$ and $\lor$ (correct because $\cpn X$ are fresh)
\smallskip \item $ %8
\forall \cpn X \; \left( \begin{aligned}
(\exists Y ( \psi[p/\top] \land \exists \cpp X \phi[\sigma]) ) %\\ 
\lor 
\exists Y \left( \psi[p/\bot] \land 
\exists \cpp X \left(\lnot \phi[\sigma] \land \bigwedge_{x \in X}(\cpp x {\leqv} \cpn x) \right) \right) 
\end{aligned} \right) 
$\\ \phantom x \hfill distribution of $\exists Y$ over $\land$ (correct because $Y \cap \freevars{\phi[\sigma]} = \emptyset$)
\smallskip \item $ %9
\forall \cpn X \; \left( \begin{aligned}
\left( \exists Y {\cup} \cpp X\, ( \psi[p/\top] \land \phi[\sigma]) \right) %\\ 
\lor 
\exists Y {\cup} \cpp X \, \left( \psi[p/\bot] \land \lnot \phi[\sigma] \land \bigwedge_{x \in X}(\cpp x {\leqv} \cpn x) \right) 
\end{aligned} \right) 
$\\ \phantom x \hfill distribution of $\exists \cpp X$ over $\land$ (correct because $\cpp X \cap \freevars \psi = \emptyset $)
\medskip \item $ %10
\forall \cpn X\;\exists Y {\cup} \cpp X \; \left( \begin{aligned}
( \psi[p/\top] \land \phi[\sigma] )  %\\ 
\lor 
\left( \psi[p/\bot] \land \lnot \phi[\sigma] \land \bigwedge_{x \in X}(\cpp x {\leqv} \cpn x) \right) 
\end{aligned} \right) 
$\\ \phantom x \hfill distribution of $\exists Y {\cup} \cpp X$ over $\lor$
\medskip \item $ %11
\forall \cpn X\;\exists Y {\cup} \cpp X \; \left( \begin{aligned}
\left( ( \psi[p/\top] \land \phi[\sigma] ) \lor
   \left( \psi[p/\bot] \land \lnot \phi[\sigma]\right) \right) %\\ 
   \land
\left(\lnot \phi[\sigma] \limp \bigwedge_{x \in X}(\cpp x {\leqv} \cpn x) \right) 
\end{aligned} \right)
$\\ \phantom x \hfill propositional calculus
\medskip \item $ %12
\forall \cpn X\;\exists Y {\cup} \cpp X \; \left( \left(
\psi[p/\phi[\sigma]]
 \land \left(\lnot \phi[\sigma] \limp \bigwedge_{x \in X}(\cpp x {\leqv} \cpn x) \right) \right) \right)
$\\ \phantom x \hfill by Fact~\ref{lemma:subst}.\ref{lemma:subst_cases}  (applies: $p$ is not in the scope of any quantifier in $\psi$, so $\phi[\sigma]$ is free for $p$ in $\psi$)
\medskip \item $ %13
\forall \cpn X\;\exists Y {\cup} \cpp X\; \left( 
\left( \psi \land 
  \bigwedge_{x \in X}\left( \lnot p \limp (\cpp x {\leqv} \cpn x) \right)
\right) [p/\phi[\sigma]] \right) 
$\\ \phantom x \hfill substitution  (correct: $\cpp X \cap \freevars\psi = \emptyset$ and $p \notin \cpn X$) %(correct because $p \notin X \cup \cpn X$)
\medskip \item $ %14
\forall \cpn X \; \exists Y {\cup} \cpp X {\cup} \{p\} \; \left( \begin{aligned}
\psi \land (p \leqv \phi[\sigma]) %\\ 
\land
  \bigwedge_{x \in X}\left( \lnot p \limp (\cpp x {\leqv} \cpn x) \right)
\end{aligned} \right)
$\\ \phantom x \hfill by Fact~\ref{lemma:subst}.\ref{lemma:subst_equiv}  (applies: 
$p$ not in the scope of any quantifier in $\psi$ and 
$p \notin \freevars{\phi[\sigma]}$) % does not occur in $\phi[\sigma]$) 
\end{enumerate}

For the second equivalence %case $Q = \forall$ 
we use that 
$\exists Y \psi[p/\forall X \phi]$ is logically equivalent to 
$\exists Y \left( (\psi[p/\lnot q])[q/\exists X \lnot \phi] \right)$ if $q$ is fresh.
Then the first case applies: the latter is logically equivalent to 
%\ah{par convention, pas de parentheses autour de $(\lnot q [\sigma])$}
$$ % \begin{align*}
\forall \cpn X\;\exists Y {\cup} \cpp X {\cup} \{q\} \; \left( \begin{aligned}
\psi[p/\lnot q] \land 
(q \leqv \lnot \phi[\sigma]) %\\ 
\land
\left( \bigwedge_{x \in X}(\lnot q \limp (\cpp x {\leqv} \cpn x)) \right) 
\end{aligned} \right) , 
$$ %\end{align*}
which is logically equivalent to 
$$\forall \cpn X\;\exists Y {\cup} \cpp X {\cup} \{p\} \; \left( \begin{aligned}
\psi \land (p \leqv \phi[\sigma]) %\\ 
\land
\left( \bigwedge_{x \in X}(\phantom{\lnot} p \limp (\cpp x {\leqv} \cpn x)) \right) 
\end{aligned} \right) . $$
\end{pf}

We are going to iterate the application of Lemma~\ref{lemma:extract_quantifier} in order to prenex all outermost quantifiers. 
First of all, we observe that we can identify all outermost quantifiers of a given full QBF~$\phi$. 

% \begin{fact}\label{lemma:boolean_combination}
% Let $\phi \in \lang{}$ be a full QBF. Then there exist 
% a boolean $\beta \in \lang{bool}$ and a substitution
% $\{p_1 / \exists X_1\phi_1,\ldots, p_k / \exists X_k\phi_k\}$ such that
% $\phi = \beta[ p_1/\exists X_1\phi_1,\ldots,p_k/\exists X_k\phi_k ]$, 
% $\lngth\phi = \lngth\beta + \sum_{i \leq k} \card{X_i} + \sum_{i \leq k} \lngth{\phi_i}$, and
% $\mdepth\phi = \max_{i\leq k}(1 + \mdepth{\phi_i})$. 
% The formula $\beta$ can be chosen such that $p_i \notin \bigcup_{j\leq k} \freevars{\phi_j}$.
% %and such that $X_i$ and $X_j$ are disjoint for $i \neq j$. 
% \end{fact}

% \begin{fact}[univ]\label{lemma:boolean_combination}
% Let $\phi \in \lang{}$ be a full QBF. Then there exist 
% a boolean $\beta \in \lang{bool}$ and a substitution
% $\{p_1 / \exists X_1\phi_1,\ldots, p_{\ell} / \exists X_{\ell}\phi_{\ell}, p_{\ell+1} / \forall X_{\ell+1} \phi_{\ell+1}, \ldots, p_{k} / \forall X_{k} \phi_{k}\}$ such that
% $\phi = \beta[ p_1/\exists X_1\phi_1,\ldots,p_{\ell}/\exists X_{\ell}\phi_{\ell}, p_{\ell+1} / \forall X_{\ell+1} \phi_{\ell+1}, \ldots, p_{k} / \forall X_{k} \phi_{k} ]$, 
% $\lngth\phi = \lngth\beta + \sum_{i \leq k} \card{X_i} + \sum_{i \leq k} \lngth{\phi_i}$, and
% $\mdepth\phi = \max_{i\leq k}(1 + \mdepth{\phi_i})$. 
% The formula $\beta$ can be chosen such that $p_i \notin \bigcup_{j\leq k} \freevars{\phi_j}$.
% %and such that $X_i$ and $X_j$ are disjoint for $i \neq j$. 
% \end{fact}

\medskip\begin{fact}\label{lemma:boolean_combination}
For every full QBF  $\phi \in \lang{}$ there exist 
a boolean formula $\beta \in \lang{bool}$ and a substitution
$\{p_1 / Q_1 X_1\phi_1,\ldots, p_k / Q_k X_k\phi_k\}$ such that
$$\phi = \beta[ p_1/Q_1 X_1\phi_1,\ldots,p_k/Q_k X_k\phi_k ] , $$ 
$\lngth\phi = \lngth\beta + \sum_{i \leq k} \card{X_i} + \sum_{i \leq k} \lngth{\phi_i}$, and
$\mdepth\phi = \max_{i\leq k}(1 + \mdepth{\phi_i})$. 
The formula $\beta$ can be chosen such that $p_i \notin \bigcup_{j\leq k} \freevars{\phi_j}$.
%and such that $X_i$ and $X_j$ are disjoint for $i \neq j$. 
\end{fact}

\medskip
The above fact also holds when $\phi$ is boolean: then $k=0$ (the substitution is empty), $\phi$ equals $\beta$, and 
$\mdepth{\phi} = 0$. % \max_{i\leq k}(1 + \mdepth{\phi_i}) = 0$. 

We are ready to define a transformation simultaneously prenexing all outermost quantifiers. 

% \begin{definition}\label{def:sf_level_one}
% Let $\phi \in \lang{}$ be a full QBF. 
% Let $\beta \in \lang{bool}$ and
% $\{p_1 / \exists X_1\phi_1,\ldots, p_k / \exists X_k\phi_k\}$ be as in Fact~\ref{lemma:boolean_combination}.
% %Let $X = \bigcup_{i \leq k} X_i$. 
% For $i \leq k$ and $x \in X_i$, let $\cpp{x_i}$ and $\cpn{x_i}$ be fresh variables.
% %Let $\cpp X = \{ \cpp x \suchthat x \in X_i, i \leq k \}$ be a set of fresh variables. 
% Let $\sigma_i = \{ x / \cpp{x_i} \suchthat x \in X_i \}$. 
% \begin{align*}
% \transa{\phi } & = 
% \left( 
% \bigwedge_{i\leq k} \left( 
% (p_i\leqv \phi_i[\sigma_i]) \land \left( \lnot p_i \limp \bigwedge_{x \in X_i}(\cpp{x_i}\leqv \cpn{x_i}) \right) \right) \right) 
% \land \beta ; 
% \\
% \transe{\phi } & = 
% \left( \bigwedge_{i\leq k} \left( 
% (p_i\leqv \phi_i[\sigma_i]) \land \left( \lnot p_i \limp \bigwedge_{x \in X_i}(\cpp{x_i}\leqv \cpn{x_i}) \right) \right) \right) 
% \limp \beta ;
% \\
% \levelzeronegcopies{\phi} & = \{ \cpn{x_i} \suchthat x \in X_i, i \leq k\} ; & %\text{a set of fresh variables} 
% \\
% \levelzeroposcopies{\phi} & = \{ \cpp{x_i} \suchthat x \in X_i, i \leq k\} \cup \{p_i\suchthat i \leq k\} . 
% \end{align*}
% \end{definition} 

\medskip\begin{definition}\label{def:sf_level_one}
Let $\phi \in \lang{}$ be a full QBF. 
Let $\beta \in \lang{bool}$ and
$\{p_1 / Q_1 X_1\phi_1,\ldots, p_k / Q_k X_k\phi_k\}$ be as in Fact~\ref{lemma:boolean_combination}.
For $i \leq k$ and $x \in X_i$, let $\cpp{x_i}$ and $\cpn{x_i}$ be fresh variables.
Let $\sigma_i = \{ x / \cpp{x_i} \suchthat x \in X_i \}$. 
\begin{align*}
\transa{\phi } & = 
\left( \begin{aligned}
\left( \bigwedge_{i\leq k} \left( p_i {\leqv} \phi_i[\sigma_i] \right) \right) %\\ 
\land 
\bigwedge_{i\leq k, Q_i = \exists} \left( 
\lnot           p_i {\limp} \bigwedge_{x \in X_i}(\cpp{x_i} {\leqv} \cpn{x_i}) \right) %\\ 
\land 
\bigwedge_{i\leq k, Q_i = \forall} \left( 
\phantom{\lnot} p_i {\limp} \bigwedge_{x \in X_i}(\cpp{x_i} {\leqv} \cpn{x_i}) \right)
\end{aligned} \right) 
\land \beta ; 
\\
\transe{\phi } & = 
\left( \begin{aligned}
\left( \bigwedge_{i\leq k} \left( p_i {\leqv} \phi_i[\sigma_i] \right) \right) %\\ 
\land 
\bigwedge_{i\leq k, Q_i = \exists} \left( 
\lnot           p_i {\limp} \bigwedge_{x \in X_i}(\cpp{x_i} {\leqv} \cpn{x_i}) \right) %\\ 
\land \!\!
\bigwedge_{i\leq k, Q_i = \forall} \left( 
\phantom{\lnot} p_i {\limp} \bigwedge_{x \in X_i}(\cpp{x_i} {\leqv} \cpn{x_i}) \right)
\end{aligned} 
\right) \limp \beta ; 
\\
\levelzeronegcopies{\phi} & = \{ \cpn{x_i} \suchthat x \in X_i, i \leq k\} ; 
\\
\levelzeroposcopies{\phi} & = \{ \cpp{x_i} \suchthat x \in X_i, i \leq k\} \cup \{p_i\suchthat i \leq k\} . 
\end{align*}
\end{definition} 

% \begin{definition}[univ2 ter]\label{def:sf_level_one}
% Let $\phi \in \lang{}$ be a full QBF. 
% Let $\beta \in \lang{bool}$ and
% $\{p_1 / Q_1 X_1\phi_1,\ldots, p_k / Q_k X_k\phi_k\}$ be as in Fact~\ref{lemma:boolean_combination}.
% For $i \leq k$ and $x \in X_i$, let $\cpp{x_i}$ and $\cpn{x_i}$ be fresh variables.
% Let $\sigma_i = \{ x / \cpp{x_i} \suchthat x \in X_i \}$.
% Let $\ell_i = \lnot p_i$ if   $Q_i = \exists$ and $\ell_i = p_i$ otherwise.
% \begin{align*}
% \transa{\phi } & = 
% \left( 
% \bigwedge_{i\leq k} \left( 
% p_i\leqv \phi_i[\sigma_i]\right) \land \left(\ell_i \limp \bigwedge_{x \in X_i}(\cpp{x_i}\leqv \cpn{x_i})  \right)\right)
% \land \beta ; 
% \\
% \transe{\phi } & = 
% \left( 
% \bigwedge_{i\leq k} \left( 
% p_i\leqv \phi_i[\sigma_i]\right) \land  \left( \ell_i \limp \bigwedge_{x \in X_i}(\cpp{x_i}\leqv \cpn{x_i})  \right)\right)
% \limp \beta ; 
% \\
% \levelzeronegcopies{\phi} & = \{ \cpn{x_i} \suchthat x \in X_i, i \leq k\} ; & %\text{a set of fresh variables} 
% \\
% \levelzeroposcopies{\phi} & = \{ \cpp{x_i} \suchthat x \in X_i, i \leq k\} \cup \{p_i\suchthat i \leq k\} . 
% \end{align*}
% \end{definition} 

The substitutions $\sigma_i$ replace the elements of $X_i$ by fresh variables and thereby ensure that in $\transa{\phi}$ and $\transe{\phi}$, none of the variables of $X_i$ is quantified. 
%\ah{avant: "none of the variables of $\levelzeronegcopies{\phi} \cup \levelzeroposcopies{\phi}$" is quantified." }

\goodbreak
\medskip\begin{example}
Let
\begin{align*}
\phi &= \exists x (\psi \land \lnot \exists x \chi) \land \lnot \forall y \rho
\\&= (p_1 \land \lnot p_2)[ p_1/\exists x (\psi \land \lnot \forall x \chi) , p_2/\forall y \rho] ,
\end{align*}
where $\psi,\chi,\rho$ are all boolean. Then 
$\levelzeronegcopies{\phi} = \{\cpn x, \cpn y\}$, 
$\levelzeroposcopies{\phi} = \{\cpp x, \cpp y, p_1, p_2\}$, and 
\begin{align*}
\transe{\phi} &= 
\left(\begin{aligned}
& (p_1 \leqv (\psi \land \lnot \exists x \chi)[x/\cpp x]) \\
\land\ & (\lnot p_1 \limp (\cpp x {\leqv} \cpn x)) \land \top \\
\land\ & (p_2 \leqv \rho[y/\cpp y])\\
\land\ & \top \land (p_2 \limp (\cpp y {\leqv} \cpn y)) 
\end{aligned}\right) \limp 
(p_1 \land \lnot p_2) .
\end{align*}
The formula
$$\exists \levelzeronegcopies{\phi} \forall \levelzeroposcopies{\phi} \transe{\phi} =
 \exists \{\cpn x, \cpn y\} \forall \{\cpp x, \cpp y, p_1, p_2\} \transe{\phi} $$ 
is logically equivalent to $\phi$, as will follow from the next lemma. 
\end{example}

% \begin{lemma}[nouvelle]\label{lemma:k_quantifiers}
% Let $\phi \in \lang{}$ be a  full QBF and let $Q$ be either $\forall$ or $\exists$. Then 
% \begin{enumerate*}
% \item $Q \levelzeronegcopies{\phi} \, \transq{Q}{\phi}$ is equivalent to $\phi$; 
% % As such, $\transa{\phi }$ and $\phi$ are equivalid, and $\transe{\phi }$ and $\phi$ are equisatisfiable.
% %\item Both $\forall \levelzeronegcopies{\phi) \transe{\phi }$ and $\exists \levelzeronegcopies{\phi) \transa{\phi}$ are equivalent to $\phi$.
% \item the length of $\transq{Q}{\phi }$ is linear in the length of $\phi$; and
% \item the quantifier depth of $\transq{Q}{\phi}$ equals that of $\phi$.
% \end{enumerate*}
% \end{lemma}

\medskip\begin{lemma}\label{lemma:k_quantifiers}
Let $\phi \in \lang{}$ be a full QBF and let $Q$ be either $\forall$ or $\exists$. Then: 
\begin{enumerate}
\item\label{item:k_equivalence}
$Q \levelzeronegcopies{\phi} Q' \levelzeroposcopies{\phi}\, \transq{Q}{\phi} \equiv \phi$, where $Q'$ is the dual of $Q$. 
\item\label{item:k_quantifiers}
If $\phi$ is boolean then $\mdepth{\transq{Q}{\phi}} = 0$, 
else $\mdepth{\transq{Q}{\phi}} = \mdepth{\phi} - 1$. 
\item 
%the length of $\transq{Q}{\phi }$ is linear in the length of $\phi$; 
$\lngth{ \transa{ \phi } } + 3 \nbvar{\transa{ \phi }} + 6 \nblock{\transa{ \phi }} $\\$ \leq \lngth{\phi} + 3 \nbvar{\phi} + 6 \nblock{\phi}$.
\end{enumerate}
\end{lemma}
\begin{pf}
Let $\phi \in \lang{}$ be a full QBF. 
Let $\beta \in \lang{bool}$ and
$\{p_1 / Q_1 X_1\phi_1,\ldots, p_k / Q_k X_k\phi_k\}$ be as in Definition~\ref{def:sf_level_one} (and Fact~\ref{lemma:boolean_combination}). 

%\begin{enumerate}
\bigskip%item
Let us prove \emph{item (1)} for the case where $Q$ is $\forall$. 
We define for all $i \leq k$:
\begin{align*}
\levelzeronegcopiesupto{\phi} i &= \{x_j^- \suchthat j \leq i \text{ and } x \in X_j\} ,
\\
\levelzeroposcopiesupto{\phi} i &= \{x_j^+ \suchthat j \leq i \text{ and } x \in X_j\} \cup \{p_j \suchthat j\leq i\} . 
\end{align*}
We show by induction on $k \geq 0$ that $\phi$ is logically equivalent to 
$\forall \levelzeronegcopiesupto{\phi} k \, 
\exists \levelzeroposcopiesupto{\phi} k \transa{\phi } $, 
that is, to
\begin{align*}
\forall \levelzeronegcopiesupto{\phi} k \; 
\exists \levelzeroposcopiesupto{\phi} k 
\left( 
\begin{aligned}
\beta \land 
\left( \bigwedge_{i\leq k} \left( p_i\leqv \phi_i[\sigma_i] \right) \right) \\ \land 
\bigwedge_{\substack{i\leq k\\ Q_i = \exists}} \left( 
\lnot           p_i \limp \bigwedge_{x \in X_i}(\cpp{x_i} {\leqv} \cpn{x_i}) \right)
\\ \land 
\bigwedge_{\substack{i\leq k\\ Q_i = \forall}} \left( 
\phantom{\lnot} p_i \limp \bigwedge_{x \in X_i}(\cpp{x_i} {\leqv} \cpn{x_i}) \right)
\end{aligned}
\right) .
\end{align*}
In the base case $k=0$ the full QBF $\phi$ is boolean. 
Hence the set of $p_i$, the set of $\cpp{x_i}$, and the set of $\cpn{x_i}$ are all empty 
and $\phi$ and $\transa\phi$ both equal $\beta$; therefore the equivalence trivially holds. 
For the induction step we give the following sequence of logically equivalent formulas: 
\goodbreak
\begin{enumerate}
\item $ %a
\phi$
\smallskip\item $ %b
\beta[ p_i/Q_i X_i\phi_i \suchthat i \leq k+1 ] 
$ \hfill by Fact~\ref{lemma:boolean_combination} 
\smallskip\item $ %c
\left( \beta[ p_i/Q_i X_i\phi_i \suchthat i \leq k ] \right) [ p_{k+1} / Q_{k+1} X_{k+1} \phi_{k+1} ]
$ \hfill by Fact~\ref{lemma:compo_subst} (applies: $p_{k+1}$ is distinct 
\\ \phantom x \hfill from $p_1,\ldots p_k$ and 
$p_{k+1} \notin \bigcup_{i \leq k} \freevars{\phi_i}$) 
\bigskip\item $ %d
\left( 
\forall \levelzeronegcopiesupto{\phi} k 
\exists \levelzeroposcopiesupto{\phi} k %\cup \{p_i\suchthat i \leq k\} \;
\left(   \begin{aligned} 
\beta \land \left( \bigwedge_{i\leq k} \left( p_i\leqv \phi_i[\sigma_i] \right) \right) \\ \land
\bigwedge_{\substack{i\leq k\\ Q_i = \exists}} \left( 
\lnot           p_i \limp \bigwedge_{x \in X_i}(\cpp{x_i} {\leqv} \cpn{x_i}) \right)
\\
 \land 
\bigwedge_{\substack{i\leq k\\ Q_i = \forall}} \left( 
\phantom{\lnot} p_i \limp \bigwedge_{x \in X_i}(\cpp{x_i} {\leqv} \cpn{x_i}) \right)
  \end{aligned}  
\right) 
\right) 
%$ \\ \phantom{xxxxxxxxxxxxxxxxxxxxxxxxxx} $
\, [ p_{k+1} / Q_{k+1} X_{k+1} \phi_{k+1} ]
$ \hfill by induction hypothesis 
\medskip\item $ %e
\forall \levelzeronegcopiesupto{\phi} k %\;\; 
\exists \levelzeroposcopiesupto{\phi} k %\; 
%$ \\ \phantom{x} $
\left( \left( \begin{aligned} 
\beta \land \left( \bigwedge_{i\leq k} (p_i\leqv \phi_i[\sigma_i]) \right) \\ \land 
\bigwedge_{\substack{i\leq k\\ Q_i = \exists}} \left( 
\lnot           p_i \limp \bigwedge_{x \in X_i}(\cpp{x_i} {\leqv} \cpn{x_i}) \right)
\\ \land 
\bigwedge_{\substack{i\leq k\\ Q_i = \forall}} \left( 
\phantom{\lnot} p_i \limp \bigwedge_{x \in X_i}(\cpp{x_i} {\leqv} \cpn{x_i}) \right)
\end{aligned}
\right)
[ p_{k+1} / Q_{k+1} X_{k+1} \phi_{k+1} ] \right)
$ \\ \phantom x \hfill because
$\levelzeronegcopiesupto{\phi} k \cup \levelzeroposcopiesupto{\phi} k$ 
and $\{p_{k+1}\} \cup \freevars{\phi_{k+1}}$ are disjoint 
($\levelzeronegcopiesupto{\phi} k \cup \levelzeroposcopiesupto{\phi} k$ being fresh) 
\medskip\item $ %f
\forall \levelzeronegcopiesupto{\phi}{k} \forall \cpn X_{k+1} \, 
\exists \levelzeroposcopiesupto{\phi}{k+1} {\cup} \cpp X_{k+1} {\cup} \{p_{k+1}\} \, 
%$ \\ \phantom{xxxxxxx} $
\left( \begin{aligned} 
\beta \land \left( \bigwedge_{i\leq k+1} (p_i\leqv \phi_i[\sigma_i])  \right) \\ \land 
\bigwedge_{\substack{i\leq k+1\\ Q_i = \exists}} \left( 
\lnot           p_i \limp \bigwedge_{x \in X_i}(\cpp{x_i} {\leqv} \cpn{x_i}) \right)
\\ \land 
\bigwedge_{\substack{i\leq k+1\\ Q_i = \forall}} \left( 
\phantom{\lnot} p_i \limp \bigwedge_{x \in X_i}(\cpp{x_i} {\leqv} \cpn{x_i}) \right)
\end{aligned} 
\right) 
$ \\ \phantom x \hfill by Lemma~\ref{lemma:extract_quantifier} (applies: the sets
$\levelzeroposcopiesupto{\phi} k$ and $\freevars{\phi_{k+1}} $ are disjoint)
\smallskip\item $ %g
\forall \levelzeronegcopiesupto{\phi} {k+1} \, 
\exists \levelzeroposcopiesupto{\phi} {k+1} \transa{\phi } . 
$
\end{enumerate}

\bigskip
The case where $Q$ is $\exists$ follows directly from the case where $Q$ is $\forall$: 
it suffices to show $\transe{\phi } \equiv \lnot \transa{\lnot \phi}$, that is, that
$\transe{\beta[ p_1/\exists X_1\phi_1,\ldots,p_k/\exists X_k\phi_k ] 
}
\equiv 
\lnot \transa{(\lnot \beta) [ p_1/\exists X_1\phi_1,\ldots,p_k/\exists X_k\phi_k ]
} .$ 

% \item
% The length of $\phi$ is 
% $ \lngth{  \phi } = 
% \lngth{\beta} + \sum_{i\leq k} (1+\card{X_i} + \lngth{\phi_i}) $; 
% for the length of $\transa{ \phi }$ we have: 
% \begin{align*}
% \lngth{ \transa{ \phi } } 
% &=
% \lngth{
% %\exists \{\cpp x \suchthat x \in X\} \cup \{p_i\suchthat i \leq k\} \;\; \left( 
% \left( 
% \bigwedge_{i\leq k} \left( 
% (p_i\leqv \phi_i[\sigma_i]) \land \left(\lnot p_i \limp \bigwedge_{x \in X_i}(\cpp{x_i}\leqv \cpn{x_i}) \right) \right) \right) 
% \land \beta %\right) 
% } 
% \\&\leq
% %1+\left( \sum_{i\leq k} (\card{X_i}+1) \right) + 
% \left( \sum_{i\leq k}(2 + \lngth{\phi_i} + 4 + 4\card{X_i}) \right) + 
% \lngth\beta 
% \\&\leq
% %1+\left( \sum_{i\leq k} \card{X_i} \right) + k +
% 4\left(\sum_{i\leq k} (1+\lngth{\phi_i} + \card{X_i})\right)+\lngth\beta
% \\&\leq 
% 4\left(\lngth\beta+ \sum_{i\leq k} (1+\lngth{\phi_i} + \card{X_i})\right)
% \\&\leq 
% 4 \lngth{ \phi} . 
% \end{align*}

% \newpage
% (SAUT DE PAGE)
% \newpage

\bigskip%item
Concerning \emph{item 2} of the lemma about quantifier depth, if $k=0$ then $\phi$ is boolean and $\mdepth{\phi} = \mdepth{ \transe{\phi} } = \mdepth{ \transa{\phi } } = 0$; otherwise
\begin{align*}
\mdepth{\phi} 
&= 1 + \max_{i\leq k}\mdepth{\phi_i} 
\\&= 1+ \mdepth{ \transe{\phi} } 
\\&= 1+ \mdepth{ \transa{\phi } } . 
\end{align*}
% \ah{
% R3: 
% "Fact 4:\\
% - "size of $\varphi$ is given as $\Vert \beta \Vert + \sum\limits_{i \leq k} (|X_i| + \Vert \varphi_i \Vert)$ (appears to be correct)\\
% - "in line 179 quoted as $\Vert \beta \Vert + \sum\limits_{i \leq k} (|X_i| + \Vert \varphi_i \Vert + 1)$\\
% - "propable reason: adding $Q_i$ increases size by 1, but removing $p_i$ decreases by 1, cancelling each other out\\
% - "leads to incorrect size estimations in Lemma 4 and proof of theorem 9\\
% - "theorem 9.3: Should still be linear, but $9* \Vert \varphi \Vert$ instead of 8

% J'ai corrigé.
% }

\bigskip%item
Finally, concerning \emph{item 3} of the lemma about formula length, let us prove that 
\begin{align*}
\lngth{ \transa{ \phi } }  
\leq 
\lngth{\phi} %&
+ 6 (\nblock{\phi} - \nblock{\transa{ \phi }}) %\\&
+ 3 (\nbvar{\phi} - \nbvar{\transa{ \phi }}) .
\end{align*}
% \ah{enlevé car ca me semble superflu : \\
% With Fact~\ref{lemma:boolean_combination}, the length of $\phi$ is 
% \begin{align*}
% \lngth{  \phi } &= 
% \lngth{\beta} %+ k 
%               + \left( \sum_{i\leq k} \card{X_i} \right) + \sum_{i\leq k} \lngth{\phi_i}
% \\&=
% \lngth{\beta} %+ \nblock{\phi} - \nblock{\transa{ \phi }} 
%               + \nbvar{\phi} - \nbvar{\transa{ \phi }} + \sum_{i\leq k} \lngth{\phi_i} , 
% \end{align*}
% }
We have: %The length of $\transa{ \phi }$ is
\goodbreak
\begin{align*}
&\lngth{ \transa{ \phi } } 
\\=\ & \lngth\beta + 1 + \lngth{ 
\bigwedge_{i\leq k} \left( p_i\leqv \phi_i[\sigma_i] \right) } %\\& 
+ \lngth{
\begin{aligned}
  \bigwedge_{\substack{i\leq k, Q_i = \exists}} \left(\lnot p_i \limp \bigwedge_{x \in X_i}(\cpp{x_i}\leqv \cpn{x_i}) \right) 
\\ \land 
  \bigwedge_{\substack{i\leq k, Q_i = \forall}} \left(\phantom{\lnot} p_i \limp \bigwedge_{x \in X_i}(\cpp{x_i}\leqv \cpn{x_i}) \right) 
\end{aligned}
} 
\\\leq\ & 
\lngth\beta + 1 + \lngth{ 
\bigwedge_{i\leq k} \left( 
p_i\leqv \phi_i[\sigma_i] \right) } %\\& 
+\lngth{
\bigwedge_{\substack{i\leq k}} \left(\lnot p_i \limp \bigwedge_{x \in X_i}(\cpp{x_i}\leqv \cpn{x_i}) \right) 
} 
\\\leq\ & 
\lngth\beta + 
\left(\sum_{i\leq k} (3 + \lngth{\phi_i} ) \right) + \left(\sum_{i \leq k} (3 + 4 \card{X_i})\right) 
\\\leq\ & 
\lngth\beta +
6k + \left(\sum_{i\leq k} \lngth{\phi_i} \right) + 
4 \left(\sum_{i \leq k} \card{X_i}\right) 
\\\leq\ & 
6k + 
3 \left(\sum_{i \leq k} \card{X_i}\right) + 
\lngth\beta + 
\left(\sum_{i \leq k} \card{X_i}\right)  + 
\left(\sum_{i\leq k} \lngth{\phi_i} \right)
\\\leq\ & 
6k + 3 \left(\sum_{i \leq k} \card{X_i}\right) + 
\lngth\phi 
\\\leq\ & 
6 (\nblock{\phi} - \nblock{\transa{\phi}}) %\\&
+ 3 (\nbvar{\phi} - \nbvar{\transa{ \phi }}) + 
\lngth\phi .
\end{align*}

The same upper bound for the length of $\transe{ \phi }$ can be shown in the same way. 
%\end{enumerate}

This ends the proof of Lemma~\ref{lemma:k_quantifiers}.
\end{pf}

\section{Prenexing All Quantifiers}\label{sec:trafo_any_level}
%%%%%%%%%%%%%%%%%%%%%%%%%%%%%%%%%

Lemma~\ref{lemma:k_quantifiers} allows us to prenex all outermost quantifiers of a full QBF. 
In order to prenex all quantifiers we iterate the transformations $\transa\phi$ and $\transe\phi$ of Definition~\ref{def:sf_level_one}, where we take care to avoid quantifier depth increase. 
\ah{reste modifié}
The latter is achieved 
by choosing the appropriate among the two equivalences of Lemma~\ref{lemma:extract_quantifier} and 
by fusing quantifier blocks. 
%At each step all outermost quantifiers are prenexed. 

% \begin{definition}[nouvelle]
% Let $Q$ be a quantifier.
% Let $\phi \in \lang{}$ be a full QBF.
% \begin{align*}
% \translationq{Q}{\phi } &= \begin{cases}
% \phi                                                             &\text{if } \phi \text{ is boolean} ; \\
% Q' X {\cup} \levelzeronegcopies{\psi} \translationq{Q'}{\psi} &\text{otherwise, for $Q'X\psi = \transq{Q}{\phi}$.} \\
% \end{cases}
% \end{align*}
% \end{definition} 

\medskip\begin{definition}
Let $Q$ be a quantifier and let $Q'$ be its dual.
Let $\phi \in \lang{}$ be a full QBF.
\begin{align*}
\translationq{Q}{\phi } &= \begin{cases}
\phi                                                    &\text{if } \phi \text{ is boolean} ; \\
Q' \levelzeroposcopies{\phi} {\cup} \levelzeronegcopies{\psi} \translationq{Q'}{\psi} &\text{otherwise, for $\psi = \transq{Q}{\phi}$.} \\
\end{cases}
\end{align*}
\end{definition} 

\medskip
Observe that the transformation is well-defined because the quantifier depth decreases in the induction step by Lemma~\ref{lemma:k_quantifiers}.\ref{item:k_quantifiers}. % (3).
The freshness of the variables $\cpp{x_i}$ substituting the $x_i$ of the subformulas $\phi_i$ according to Definition~\ref{def:sf_level_one} ensures that the translation behaves well. 

\medskip\begin{example}
Let 
\begin{align*}
\phi &= \exists x (\psi \land \lnot \exists x \chi) \land \lnot \forall y \rho
\\&= (p_1 \land \lnot p_2)[ p_1/\exists x (\psi \land \lnot \exists x \chi) , p_2/\forall y \rho] ,
\end{align*}
where $\psi$, $\chi$, and $\rho$ are all boolean. 
First:
\begin{align*}
\transe{\phi} &= 
\left(\begin{aligned}
& \phantom{\lnot} p_1 \leqv (\psi[x/\cpp x] \land \lnot \exists x \chi) \\
\land\ & \lnot p_1 \limp (\cpp x \leqv \cpn x) \\
\land\ & \phantom{\lnot} p_2 \leqv \rho[y/\cpp y]\\
\land\ & p_2 \limp (\cpp y \leqv \cpn y)
\end{aligned}\right) \limp 
(p_1 \land \lnot p_2) ,
\\
\levelzeronegcopies{\phi} &= \{\cpn x, \cpn y\} , \\ 
\levelzeroposcopies{\phi} &= \{\cpp x, \cpp y, p_1, p_2\} . 
\end{align*}
Second:
\begin{align*}
\transa{\transe{\phi}} &= 
\left(\begin{aligned}
& \phantom{\lnot}p_3 \leqv \chi[x/\cpp z]\\
\land\ & \lnot p_3 \limp (\cpp z \leqv \cpn z) \end{aligned}\right) \land %\\&\phantom{=\ }
\left(\left(\begin{aligned} & \phantom{\lnot}p_1 \leqv (\psi[x/\cpp x] \land \lnot p_3) \\ 
\land\ & \lnot p_1 \limp (\cpp x \leqv \cpn x)  \\
\land\ & \phantom{\lnot} p_2 \leqv \rho[y/\cpp y]\\
\land\ & \phantom{\lnot} p_2 \limp (\cpp y \leqv \cpn y)
\end{aligned}\right) \limp 
(p_1 \land \lnot p_2) \right) ,
\\
\levelzeronegcopies{\transe{\phi}} &= \{\cpn z\} , \\ 
\levelzeroposcopies{\transe{\phi}} &= \{\cpp z, p_3\} . 
\end{align*}
Finally, 
$\levelzeronegcopies{\transa{\transe{\phi}}} = \emptyset $
and:
\begin{align*}
\translationq{\exists}{\phi} 
&= \forall \{\cpp x, \cpp y, p_1, p_2, \cpn z \}\translationq{\forall}{\transe{\phi}}\\
&= \forall \{\cpp x, \cpp y, p_1, p_2, \cpn z \}\exists \{\cpp z, p_3\} \translationq{\exists}{\transa{\transe{\phi}}} 
\\
&= \forall \{\cpp x, \cpp y, p_1, p_2, \cpn z \}\exists \{\cpp z, p_3\} \transa{\transe{\phi}}\\
&= \forall \{\cpp x, \cpp y, p_1, p_2, \cpn z \}\exists \{\cpp z, p_3\}
%\\& \qquad
\left(\begin{aligned}&
\begin{aligned} &p_3 \leqv \chi[x/\cpp z] \\ \land\ &\lnot p_3 \limp (\cpp z \leqv \cpn z) \end{aligned}
\\& \land
\left(\left(\begin{aligned} & \phantom{\lnot}p_1 \leqv (\psi[x/\cpp x] \land \lnot p_3)\\ \land\ &
\lnot p_1 \limp (\cpp x \leqv \cpn x) \\
\land\ & \phantom{\lnot} p_2 \leqv \rho[y/\cpp y]\\
\land\ & \phantom{\lnot}  p_2 \limp (\cpp y \leqv \cpn y) 
\end{aligned}\right) 
\limp (p_1 \land \lnot p_2) \right)
\end{aligned}\right) .
% \hspace{-25mm}
% \\
% \levelzeronegcopies{\transa{\transe{\phi}}} &= \emptyset . 
\end{align*}
The formulas $\translationq{\exists}{\phi }$ and $\phi$ are equisatisfiable, as will follow from Theorem~\ref{thm:translation-is-good} below. 
\end{example}\smallskip 

% \begin{definition}
% Let $\phi \in \lang{}$ be a full QBF.
% Let $\beta \in \lang{bool}$ and
% $\{p_1 / \exists X_1\phi_1,\ldots, p_k / \exists X_k\phi_k\}$ be as in Fact~\ref{lemma:boolean_combination}.
% \begin{align*}
% \translationmc{\phi} = \beta[ p_1/\exists X_1 {\cup} \levelzeronegcopies{\phi_1}\translationq{\exists}{\phi_1},\ldots,p_k/\exists X_k {\cup} \levelzeronegcopies{\phi_k} \translationq{\exists}{\phi_k} ] . 
% \end{align*}
% \end{definition}

\medskip\begin{definition}
Let $\phi \in \lang{}$ be a full QBF.
Let $\beta \in \lang{bool}$ and
$\{p_1 / Q_1 X_1\phi_1,\ldots, p_k / Q_k X_k\phi_k\}$ be as in Fact~\ref{lemma:boolean_combination}.
\begin{align*}
\translationmc{\phi} = 
\beta[ p_1/Q_1 X_1 {\cup} \levelzeronegcopies{\phi_1}\translationq{Q_1}{\phi_1},\ldots,p_k/Q_k X_k {\cup} \levelzeronegcopies{\phi_k} \translationq{Q_k}{\phi_k} ] . 
\end{align*}
\end{definition}

\begin{example}
Let us compute $\translationmc{\phi}$ step-by-step for the above
$\phi = \exists x \phi_1 \land \lnot \forall y \rho$ 
with $\phi_1 = \psi \land \lnot \exists x \chi$ and
$\psi,\chi,\rho$ boolean. First:
%$\phi = \exists x (\psi \land \lnot \exists x \chi) \land \lnot \forall y \rho$. 
\begin{align*}
% \phi &= \exists x \phi_1 \land \lnot \forall y \rho \quad \text{where } \phi_1 = \psi \land \lnot \exists x \chi ;
% \\
\transe{\phi_1} &= 
\left(\begin{aligned}
& \phantom{\lnot} p_3 \leqv \chi[x/\cpp z] \\
\land\ & \lnot p_3 \limp (\cpp z \leqv \cpn z)
\end{aligned}\right) \limp 
(\psi \land \lnot p_3) , 
\\
\levelzeroposcopies{\phi_1} &= \{\cpp z, p_3 \} , \\
\levelzeronegcopies{\phi_1} &= \{\cpn z \} ; 
\\
%%%%%%%%%%%%%
%%%%%%%%%%%%%
\translationq{\exists}{\phi_1} 
&= \forall \{\cpp z, p_3 \}\translationq{\forall}{\transe{\phi_1}}\\
&= \forall \{\cpp z, p_3 \}\left(\left(\begin{aligned}
& \phantom{\lnot} p_3 \leqv  \chi[x/\cpp z] \\
\land\ & \lnot p_3 \limp (\cpp z \leqv \cpn z)
\end{aligned}\right) \limp 
(\psi \land \lnot p_3) \right) .
\end{align*}
Then $\levelzeronegcopies{\transa{\transe{\phi_1}}} = \emptyset$, and  
\begin{align*}
%%%%%%%%%%%%%
%%%%%%%%%%%%%
%\translationq{\exists}{\lnot \rho} 
%&= \lnot \rho\\
%%%%%%%%%%%%%
%%%%%%%%%%%%%
\translationmc{\phi} 
=\ & \left(\exists \{x\} \cup \levelzeronegcopies{\phi_1} \translationq{\exists}{\phi_1}  \right) \land \left( \lnot \forall \{y\} \cup \levelzeronegcopies{\rho} \translationq{\exists}{\rho} \right)\\
=\ & \left(\exists \{x, \cpn z\} \forall \{\cpp z, p_3 \}\left(\left(\begin{aligned}
& \phantom{\lnot} p_3 {\leqv} \chi[x/\cpp z] \\
\land\ & \lnot p_3 {\limp} (\cpp z \leqv \cpn z)
\end{aligned}\right) \limp 
(\psi {\land} \lnot p_3) \right)\right) \land 
\lnot \forall y \rho .
\end{align*}
\end{example}

\medskip\begin{theorem}
\label{thm:translation-is-good}
Let $\phi \in \lang{}$ be a full QBF. The formulas
$\phi$ and $\translationq{\exists}{\phi}$ are equisatisfiable,
$\phi$ and $\translationq{\forall}{\phi}$ are equivalid, and 
$\phi$ and $\translationmc{\phi}$ are logically equivalent.
\\
Furthermore, for $Q$ being either $\exists$ or $\forall$, we have:
\begin{enumerate}
\item 
$\translationq{Q}{\phi}$ is a prenex formula and $\translationmc{\phi}$ is a boolean combination of prenex formulas.
\item 
$\mdepth{\translationq{Q}{\phi}} = \mdepth{\translationmc{\phi}} = \mdepth{\phi} $. 
\item 
The lengths of $\translationq{Q}{\phi}$ and $\translationmc{\phi}$ are linear in the length of $\phi$. 
\end{enumerate}
\end{theorem}
\begin{pf}
The first statement follows from a slightly stronger result (*):
For any full QBF $\phi$ of quantifier depth $n \geq 0$, and any $Q \in \{\exists, \forall\}$, the formulas $\phi$ and $Q \levelzeronegcopies{\phi} \translationq{Q}{\phi}$  are logically equivalent.
We prove (*) by induction on $n$.
%by induction on the quantifier depth of formulas. % that $\phi$ and $Q \levelzeronegcopies{\phi} \translationq{Q}{\phi}$  are equivalent.
For the base case $n$ is zero and any formula $\phi$ of quantifier depth $0$ is boolean, so $\levelzeronegcopies{\phi} = \emptyset$ and $\translationq{Q}{\phi} = \phi$, and the equivalence is immediate.
For the induction step, assume the induction hypothesis holds for $n$. % and let us prove that it holds for $n+1$.
Let $\phi$ be a full QBF of quantifier depth $n+1$.
Let $Q \in \{\exists, \forall\}$ and let $Q'$ be its dual quantifier. 
%We apply the induction hypothesis on $\psi$ and $Q'$ and obtain that $\psi$ and $Q' \levelzeronegcopies{\psi} \translationq{Q'} \psi$ are equivalent.
\begin{align*}
Q \levelzeronegcopies{\phi} \translationq{Q}{\phi} &\equiv Q  \levelzeronegcopies{\phi} Q'  \levelzeroposcopies{\phi} {\cup} \levelzeronegcopies{\psi} \, \translationq{Q'}{\psi}, \text{ for } \psi = \transq{Q}{\phi}
\\&
\equiv Q  \levelzeronegcopies{\phi} Q' \levelzeroposcopies{\phi} Q' \levelzeronegcopies{\psi} \, \translationq{Q'}{\psi} 
\\&
\equiv Q  \levelzeronegcopies{\phi} Q' \levelzeroposcopies{\phi} \, \psi 
\tag{by induction hypothesis}
\\&
\equiv Q  \levelzeronegcopies{\phi} Q' \levelzeroposcopies{\phi} \, \transq{Q}{\phi} 
&\tag{by definition of $\psi$}
\\&
\equiv \phi 
&\tag{by Lemma~\ref{lemma:k_quantifiers}.\ref{item:k_equivalence}}
\end{align*}
Observe that the induction hypothesis applies because by Lemma~\ref{lemma:k_quantifiers}.\ref{item:k_quantifiers} the quantifier depth of $\transq{Q}{\phi}$ is $n$.
%Equivalence of $\phi$, $\translation{ \exists \emptyset \phi }$ and $ \translation{ \forall \emptyset \phi }$ is guaranteed by Lemma~\ref{lemma:k_quantifiers}. 
From (*) it directly follows that for any full QBF $\phi$,
$\phi$ and $\translationq{\exists}{\phi}$ are equisatisfiable,
$\phi$ and $\translationq{\forall}{\phi}$ are equivalid, and 
$\phi$ and $\translationmc{\phi}$ are logically equivalent.

The proof of the three items also proceeds by induction on quantifier depth: 
\begin{enumerate}
\item 
Let us prove that $\translationq Q \phi$ is in prenex form. 
The base case $\mdepth{\phi}= 0$ is trivial: $\phi$ is boolean. 
For the induction step suppose $\mdepth\phi \geq 1$. 
Then $\translationq Q \phi = 
Q' \levelzeroposcopies{\phi} {\cup} \levelzeronegcopies{\psi} \translationq{Q'}{\psi}$, 
for $\psi$ being $\transq{Q}{\phi}$. 
The induction hypothesis applies because $\mdepth\psi = \mdepth{ \transq{Q}{\phi} } = \mdepth\phi -1$ thanks to Lemma~\ref{lemma:k_quantifiers}.\ref{item:k_equivalence}; that is, $\translationq{Q'}{\psi}$ is in prenex form. 
Hence $\translationq Q \phi = 
Q' \levelzeroposcopies{\phi} {\cup} \levelzeronegcopies{\psi} \translationq{Q'}{\psi}$ 
is so, too. 

The formula $\translationmc{\phi}$ is a boolean combination of formulas in prenex form because each $\translationq \exists {\phi_i}$ is in prenex form. 

\item 
%We prove that $\mdepth{\translationq{Q}{\phi}} = \mdepth{\translationmc{\phi}} = \mdepth{\phi}$. 
The base case $\mdepth{\phi} = 0$ is trivial because then $\translationq{Q}{\phi} = \translationmc{\phi} = \phi$. 
When $\mdepth{\phi} \geq 1$ we have:
%For the induction step we have:
\begin{align*}
\mdepth{\translationq{Q}{\phi}} 
&= 
\mdepth{Q' \levelzeroposcopies{\phi} {\cup} \levelzeronegcopies{\psi} \, \translationq{Q'}{\psi} } , ~ \text{ for } \psi = \transq{Q}{\phi}
\\&=
1 + \mdepth{ \translationq{Q'}{\psi} } 
\\&=
1 + \mdepth{ \psi } 
&\tag{by induction hypothesis}
\\&=
\mdepth{\phi} .
\end{align*}
The induction hypothesis applies because Lemma~\ref{lemma:k_quantifiers}.\ref{item:k_quantifiers} tells us that $\mdepth{\transq{Q}{\psi}} = \mdepth\psi -1$ when $\mdepth\psi \geq 1$.

\item 
%For the base case $\phi$ is boolean and the inequality trivially holds. 
%.....
%According to the proof of Lemma~\ref{lemma:k_quantifiers}.\ref{item:k_quantifiers}, the length of $\transq{Q}{\phi}$ is bound by $4 \lngth\phi$. 
%.....
We establish by induction on quantifier depth that 
\begin{align*}
\card{\levelzeronegcopies{\phi}} + \lngth{\translationq{Q}{\phi}}  
\leq \mdepth{\phi} + 7 \nblock{\phi} + 5 \nbvar{\phi} + \lngth{\phi} .
\end{align*}
The base case is immediate.
For the induction step, let $\psi = \transq{Q}{\phi}$. 
First of all, we observe that
$\card{\levelzeronegcopies{\phi}} = \nbvar{\phi} - \nbvar{\psi}$ 
and 
$\card{\levelzeroposcopies{\phi}} = \nbvar{\phi}-\nbvar{\psi} + \nblock{\phi} - \nblock{\psi}$, 
and thus
\begin{align*}\tag{**}\label{eq:block_and_qvar}
\card{\levelzeronegcopies{\phi}} + \card{\levelzeroposcopies{\phi}} + \nblock{\psi} + 2\nbvar{\psi} 
= \nblock{\phi} + 2 \nbvar{\phi} .
\end{align*}
%
%$\card{\levelzeroposcopies{\phi}} + \nblock{\psi} + \nbvar{\psi} - \card{\levelzeronegcopies{\phi}} = \nblock{\phi} + \nbvar{\phi}$
%
%$\card{\levelzeroposcopies{\phi}} = \nblock{\phi} - \nblock{\psi} + \nbvar{\phi}-\nbvar{\phi} - \nbvar{\psi} = \card{\levelzeronegcopies{\phi}}$.
We therefore have: % assume $\lngth{\translationq{Q}{\psi}} \leq \mdepth{\psi} + 6 \nblock{\psi} + 5 \nbvar{\psi} - \card{\levelzeronegcopies{\psi}} + \lngth{\psi}$
\begin{align*}
&%\phantom{=~~ } 
\card{\levelzeronegcopies{\phi}} + \lngth{\translationq{Q}{\phi}} 
\\
=\ & 
\card{\levelzeronegcopies{\phi}} + 1 + \card{\levelzeroposcopies{\phi}} + \card{\levelzeronegcopies{\psi}} + \lngth{\translationq{Q'}{\psi}}
\\ 
\leq\ &
\card{\levelzeronegcopies{\phi}} + 1 + \card{\levelzeroposcopies{\phi}} + \mdepth{\psi} %\\& 
+ 7 \nblock{\psi} + 5 \nbvar{\psi} + \lngth{\psi}
\tag{by induction hypothesis} \\
%\\ \shortintertext{~\hfill (by induction hypothesis)} 
\leq\ & 
1+\mdepth{\psi} + \card{\levelzeronegcopies{\phi}} + \card{\levelzeroposcopies{\phi}} + \nblock{\psi} + 2 \nbvar{\psi} %\\& 
+ 6 \nblock{\psi} + 3 \nbvar{\psi} + \lngth{\psi}
\tag{by rearranging terms} \\ 
%\\ \shortintertext{~\hfill (by rearranging terms)} 
%&\text{rearrange}
\leq\ & 
\mdepth{\phi} + \card{\levelzeronegcopies{\phi}} + \card{\levelzeroposcopies{\phi}} + \nblock{\psi} + 2 \nbvar{\psi} %\\& 
+ 6 \nblock{\phi} + 3 \nbvar{\phi} + \lngth{\phi}
\tag{by Lemma~\ref{lemma:k_quantifiers}.\ref{item:k_quantifiers}} \\ 
%\\ \shortintertext{~\hfill (by Lemma~\ref{lemma:k_quantifiers}.\ref{item:k_quantifiers})} 
\leq\ & 
\mdepth{\phi} + \nblock{\phi} + 2\nbvar{\phi} %\\& 
+ 6 \nblock{\phi} + 3 \nbvar{\phi} + \lngth{\phi} 
\tag{by equality (\ref{eq:block_and_qvar})} \\
\leq\ & 
\mdepth{\phi} + 7 \nblock{\phi} + 5 \nbvar{\phi} + \lngth{\phi}
\tag{by rearranging terms} 
\end{align*}
Keeping in mind that $\mdepth{\phi} \leq \lngth\phi$ and that $\nblock{\phi} + \nbvar{\phi} \leq \lngth{\phi}$, we have 
$$ \card{\levelzeronegcopies{\phi}} + \lngth{\translationq{Q}{\phi}} \leq 9 \lngth\phi . $$
Hence the lengths of  $\translationq{\exists}{\phi}$, $\translationq{\forall}{\phi}$, and $\translationmc{\phi}$ are all linear in the length of $\phi$. 
\end{enumerate}
This ends the proof of the theorem.
\end{pf}

For the fragment of full QBFs of quantifier depth less or equal $k$, we establish the lower bound of model checking via an existing reduction of $\Sigma_{k-1}^{\ptime}$-complete problems to $\Theta_k^{\ptime}$ problems. 
The latter is the class of problems that can be solved 
%the class of languages recognizable 
by deterministic Turing machines in polynomial time with at most logarithmic many calls to a $\Sigma_{k-1}^{\ptime}$ oracle. 

% \begin{theorem}[Based on Theorem 3.2 in \cite{LukasiewiczM2017}]
% Let $A_1$ and $A_2$ be two (not necessarily distinct) $\Sigma_{k-1}^{\ptime}$-complete problems. Then, for a given pair of sets $X = \{x_1, \dots, x_n \}$ and
% $Y = \{ y_1, \dots, y_m \}$ of instances of $A_1$ and $A_2$, respectively, deciding whether $\card{\{x_i \suchthat \chi_{A_1}(x_i) = 1\}} > \card{\{ y_j \suchthat \chi_{A_2}(y_j) = 1\}}$ is $\Theta_k^{\ptime}$-hard.
% The hardness holds even if sets $X$ and $Y$ are assumed to be such that $\chi_{A_1}(x_1) \geq \dots \geq \chi_{A_1}(x_n)$ and $\chi_{A_2}(y_1) \geq \dots \geq \chi_{A_2}(y_m)$, and
% $n = m$.
% \end{theorem}

% \begin{lemma}[Immediate from Theorem 3.1 in \cite{LukasiewiczM2017}] 
% Let $A = \{\exists X_1 \phi_1, \dots, \exists X_n \phi_n \}$ and
% $B = \{ \exists Y_1 \psi_1, \dots, \exists Y_n \psi_n \}$ be sets of QBFs in prenex form of depth $k$ such that 
% $\exists X_i \phi_{i} \limp \exists X_{i+1} \phi_{i+1}$ and $\exists Y_i \psi_{i} \limp \exists Y_{i+1} \psi_{i+1}$ are valid, for $1 \leq i < n$. 
% Deciding whether there is an $i \leq n$ such that   
% $ \exists X_i \phi_i$ is satisfiable and 
% $\exists Y_i \psi_i$ is unsatisfiable 
% %$\max ( \{i \suchthat 1 \leq i \leq n, \exists X_i \phi_i \text{ is satisfiable} \} ) > \max ( \{i \suchthat 1 \leq i \leq n, \exists Y_i \psi_i \text{ is satisfiable} \} ) $
% is in $\Theta_{k+1}^{\ptime}$.
% \end{lemma}

\medskip\begin{lemma}[Immediate from Theorem 3.2 in \cite{LukasiewiczM2017}]\label{lemma:Theta}
Let $A = \{\phi_1, \dots, \phi_n \}$ and
$B = \{ \psi_1, \dots, \psi_n \}$ be sets of fully quantified QBFs in prenex form starting with an $\exists$ quantifier and of depth $k$ such that $\phi_{i} \limp \phi_{i+1}$ and $\psi_{i} \limp \psi_{i+1}$ are valid, for $1 \leq i < n$.
Deciding whether there exists $i\leq n$ such that $\phi_i$ is satisfiable and 
$\psi_i$ is unsatisfiable is $\Theta_{k+1}^{\ptime}$-hard.
\end{lemma}

\medskip\begin{corollary}\label{cor:complexity}
The satisfiability, validity, and model checking of full QBFs are \pspace-complete problems.
For the fragment of  formulas of quantifier depth less or equal $k$, for $k \geq 0$:
the satisfiability problem is $\Sigma_{k+1}^{\ptime}$-complete;
the validity problem is $\Pi_{k+1}^{\ptime}$-complete; and 
the model checking problem is $\Theta_{k+1}^{\ptime}$-complete. %hard and in $\ptime_{\parallel}^{\Sigma_{k}^{\ptime}}$.
\end{corollary}
\begin{pf}
For each of the first three classes \pspace, $\Sigma_{k+1}^{\ptime}$, and $\Pi_{k+1}^{\ptime}$, the hardness result follows directly from the corresponding results for QBFs in prenex form, and the membership result follows from Theorem~\ref{thm:translation-is-good}. 
It remains to establish $\Theta_{k+1}^{\ptime}$-completeness of model checking formulas $\phi$ of quantifier depth less or equal $k$ in a valuation $\val$. 

The $\Theta_{k+1}^{\ptime}$-hardness result follows from Lemma~\ref{lemma:Theta}: let 
$A = \{\phi_1, \dots, \phi_n \}$ and
$B = \{\psi_1, \dots, \psi_n \}$ 
be sets of fully quantified QBFs in prenex form starting with an $\exists$ quantifier and of depth $k$ 
such that $\phi_{i} \limp \phi_{i+1}$ and $\psi_{i} \limp \psi_{i+1}$ are valid, for $1 \leq i < n$. 
There is an $i \leq n$ such that $\phi_i$ is satisfiable and $\psi_i$ is unsatisfiable if and only if 
%$\emptyset \models \chi$, for 
%\begin{align*}
%F =& (\phi_1 \limp \psi_1) \limp \left( (\phi_2 \limp \psi_2) \limp \left(\dots \limp \left((\phi_n \limp \psi_n) \limp \bot\right) \right)\right)
%&\land \lnot\left((\psi_1 \limp \phi_1) \limp \left( (\psi_2 \limp \phi_2) \limp \left(\dots \limp \left((\psi_n \limp \phi_n) \limp \bot\right) \right)\right)\right)
%\end{align*}
$$ \intFml{
\bigvee_{i \leq n} \Big(\phi_i \land \lnot \psi_i\Big) 
} = 1 , $$
for some arbitrary valuation $\val$.
The latter is a model checking problem for formulas of quantifier depth $k$. 

We establish the $\Theta_{k+1}^{\ptime}$-membership result by using that $\Theta_{k+1}^{\ptime}$ equals  $\ptime_{\parallel}^{\Sigma_{k}^{\ptime}}$ 
(cf.\ Theorem 8.1 in \cite{Wagner90} and the remark following it). 
The latter is the class of problems that can be solved by deterministic Turing machines in polynomial time with at most polynomially many non-adaptive parallel calls to a $\Sigma_{k}^{\ptime}$ oracle. 
%model checking formulas of quantifier depth less or equal $k$ can be 
%$\ptime_{\parallel}^{\Sigma_{k}^{\ptime}}$ can be 
%decided in polynomial time with a $\Sigma_{k}^{\ptime}$ oracle. 
Given a valuation $\val$ and a formula $\phi$ of quantifier depth $k$, 
we first replace the free variables of $\phi$ by either $\top$ if $p \in \val$ or $\bot$ if $p \notin \val$, resulting in a formula of the form
$$\phi' = \beta[ p_1/Q_1 X_1\phi_1,\ldots,p_k/Q_k X_k\phi_k ]$$
for some substitution
$\{p_1 / Q_1 X_1\phi_1,\ldots, p_k / Q_k X_k\phi_k\}$;
we then compute 
\begin{align*}
&\translationmc{\phi'} = \beta[ p_1/Q_1 X_1 {\cup} \levelzeronegcopies{\phi_1}\translationq{Q_1}{\phi_1},\ldots, 
p_k/Q_k X_k {\cup} \levelzeronegcopies{\phi_k} \translationq{Q_k}{\phi_k} ] ; 
\end{align*}
finally we evaluate $\translationmc{\phi'}$ in $\val$, calling a $\Sigma_{k}^{\ptime}$ oracle for each of the subformulas 
$Q_i X_i {\cup} \levelzeronegcopies{\phi_i}\translationq{Q_i}{\phi_i}$. 
%
%, assume $\phi$ is cleansed and of quantifier depth $k$. Let $\translationmc{\phi} \equiv \beta[ p_1/P_1 \phi_1,\ldots,p_{n}/P_{n} \phi_{n} ]$ with $\beta, \phi_1,\dots,\phi_{n}$ are all boolean, and $P_1,\dots P_{n}$ are prefixes of depth $k$ starting with an existential quantifier. Let $P$ be the merging of $P_1$, with $P_2$, \dots, $P_{n}$, and $\exists C$. Consider the following sets: 
%$F_i = P \text{count}(n+1-i,C,\phi_1,\dots,\phi_n)$ 
%$G_i = P \text{count}(n+1-i,C,\phi_1,\dots,\phi_n) \land \beta[ p_1/ \phi_1,\ldots,p_{n}/ \phi_{n} ]$
%where $\exists C \text{count}(i,C,x_1,\dots,x_n)$ is a counter encoding, satisfied iff at least $i$ of $x_1,\dots,x_n$ are true (and using $C$ as auxiliary helper variables). For example Sinz's counter~\cite{Sinz2005} can be defined as the set of clauses
\end{pf}

\section{Discussion: Implementation and Evaluation}\label{sec:discu}
%%%%%%%%%%%%%%%%%%%%%%%%%%%%%%%%%

A straightforward implementation of our transformation could take the form of a ``prenexing $\Rightarrow$ optional preprocessing $\Rightarrow$ solving'' pipeline,
where the preprocessing stage refers to standard techniques such as blocked-clause elimination and unit-propagation. 
One avenue for experimentation would then be to contrast the time needed by the different stages. 
However, compared to the other two stages, the task of prenexing with our method is computationally very cheap. 
As such, the experiments may tell the reader how good of a parser we wrote and how carefully we optimised some low-level code, but not much about prenexing itself.

A more appealing avenue for experimentation would be to compare an implementation of our prenexing approach to an alternative strategy (or to solving without prenexing). The main obstacle here is that the formulas/circuits and the tools we found are unsuitable, typically because they are restricted to prenex input. 
As to benchmarks, the QBFEVAL / QBFGallery collections are either prenex (QCIR 2023 and QCIR 2020) or contain only monotone operators (pre-2020 benchmarks in Boole format), in which case preexisting standard techniques apply just fine.\footnote{
\url{https://qbf23.pages.sai.jku.at/gallery/}
}
Some more benchmarks are described in \cite{DBLP:conf/icla/StephanM09}, but it seems that they are no longer available.\footnote{
\url{http://www.info.univ-angers.fr/pub/damota/qbf} is a dead link. }
As to software, we have checked format converters with QCIR input: 
Klieber's qcir-to-qdimacs.py\footnote{\url{https://www.wklieber.com/ghostq/qcir-converter.html} }; 
Tentrup's qcir2aiger\footnote{\url{https://github.com/ltentrup/quabs} };
%, qcir2aiger est un traducteur qui vient avec Quabs que l'on mentionne par ailleurs (et pour lequel tu donnes déjà un lien).
as well as solvers with QCIR input: 
Janota's cquesto\footnote{\url{https://github.com/MikolasJanota/qesto}} and 
qfun\footnote{\url{https://github.com/MikolasJanota/qfun}}; 
Slivosky's Qute\footnote{\url{https://github.com/fslivovsky/qute/}}; 
Klieber's post-2010 GhostQ\footnote{
\url{https://www.wklieber.com/ghostq/}
%"The above versions of GhostQ no longer support non-prenex input (except the old 2010 version)."
}; 
Tentrup's Quabs\footnote{\url{https://finkbeiner.groups.cispa.de/tools/quabs/index.html}}; 
all of them assumed prenexed input. 
(We mention only one author for the sake of brevity.)

Actually we seem here to have hit a chicken-and-egg problem: if preprocessors and solvers target already prenexed formats, then someone who wants to leverage QBF technology to solve their problem has to write their instance in prenex form. If instances, benchmarks, and competitions use prenex form, and if no good prenexing algorithms are known, then the incentive to develop and maintain QBF tools able to handle non-prenex input remains small. 
We contribute to breaking this cycle by providing an easy-to-implement prenexing algorithm with optimal asymptotic performance.

Given the scarcity of relevant benchmarks and software, the alternative would be to contrast an implementation of our proposed approach with that of a naive prenexer that we would implement ourselves on benchmarks that we also create ourselves and show experimentally that our proposed approach scales better than a naive prenexer that incurs an exponential blow-up. Such results would not increase our own confidence (let alone the confidence of an independent reader) in the merit of the proposed approach beyond what is already shown by our theoretical results, and they may even seem contrived.

Our transformation applies to formulas represented as strings, as opposed to a representation by directed acyclic graphs (DAGs). 
The latter allow structure sharing and thereby a more compact representation. 
We leave a version of our transformation for DAGs to future work. 

An example is the full version of the QCIR format, which is a non-prenex, non-CNF format for QBF solvers. 
Full QCIR covers full QBF, but not the other way round. However, full QBF does not cover QCIR because the latter allows ``structure sharing'': it is not clear how a circuit/DAG (QCIR) can be converted to a formula/tree (full QBF) without incurring either an exponential blow-up in the size or an increase in the quantifier depth. 
We elected to focus the presentation on formulas rather than circuits in this paper because prenexing formulas seemed to be a more elementary problem than prenexing circuits. 
%Nevertheless, our technique can be adapted to convert full QCIR to prenex QCIR under the same size guarantees. 
We believe that our technique can be adapted in order to convert full QCIR to prenex QCIR under the same size guarantees; 
the details however remain to be elaborated and are left to future work. 
%\abda{corrigé les quptes ``...'' au lieu de "..." et remplacé un rich par full.}
%\abda{On devrait peut-être être juste dire qu'on a focus sur les formules parce que c'est fondamental et on laisse adapter aux circuits à future work. Pour le rebuttal, ça ne me dérangeait pas d'être sûrs de nous, mais ici il vaut mieux être plus prudent: Les transformations évidentes, on sait maintenant qu'elles demandent parfois plusieurs années.}

%%%%%%%%%%%%%%%%%%%%%%%%%%%%%%%%%
\section{Conclusion}\label{sec:conclu}
%%%%%%%%%%%%%%%%%%%%%%%%%%%%%%%%%

We have introduced a prenexing transformation for the full language of quantified boolean formulas that is polynomial and preserves quantifier depth. 
As far as we know we are the first to improve the known exponential reduction. 
This has allowed us to derive complexity results for satisfiability, validity, and model checking. 
%These bounds are tight in all cases but for model checking the class of formulas of quantifier depth $k$. We conjecture that the lower bound is tight. For the case $k=1$ this follows from the identity of $\Theta_{2}^{\ptime}$ and $\ptime_{\parallel}^{\Sigma_{1}^{\ptime}}$ \cite[Theorem 7]{DBLP:journals/iandc/BussH91}. 
In particular, we have provided a new complexity result for the polynomial hierarchy of full QBFs. 
Indeed, while it is immediate to adapt the textbook proofs of \pspace membership from prenex QBFs to full QBFs, things get more subtle when looking at bounded depth formulas. 
Our Corollary~\ref{cor:complexity} that model checking full QBFs of depth $k$ is $\Theta_{k+1}^{\ptime}$-hard contrasts with the known $D_k^\ptime$ upper bound of model checking prenex QBFs. 
It follows that any proof of a tight upper-bound on the model checking of prenex QBFs would be too strong to be applicable to full QBFs, 
unless the polynomial hierarchy collapses to the $k$-th level 
(which would be a consequence of showing a $\Theta_{k+1}^{\ptime}$-hard problem to be in $D_k^\ptime$)~\cite{HemaspaandraHH1998}.

While these results are theoretical, they can be expected to have practical impact whenever applications lead to formulas with arbitrarily nested quantifiers and boolean operators. 

%It would clearly be interesting to establish membership of the model checking problem for formulas of quantifier depth $k$ in $\Theta_{k+1}^{\ptime}$. 

%boolean circuits ... 

%\section*{Acknowledgment}
%[Put sponsor acknowledgments in the unnumbered footnote on the first page.]

%\section*{References}
%Unless there are six authors or more give all authors' names; do not use ``et al.''. Papers that have not been published, even if they have been submitted for publication, should be cited as ``unpublished'' \cite{b4}. Papers that have been accepted for publication should be cited as ``in press'' \cite{b5}. Capitalize only the first word in a paper title, except for proper nouns and element symbols.

%\bibliographystyle{plainurl} %{apalike}
\bibliography{references}
\end{document}

%% file: preamble.tex
\usepackage{amsmath}
\usepackage{mathtools}
\usepackage{txfonts} % cf. command `assign'
\usepackage{xspace}
\usepackage{xcolor}
\usepackage[utf8]{inputenc}
\usepackage{hyperref}
\usepackage{booktabs}
\usepackage{tikz}
\usepackage{comment}

\excludecomment{shortproofs}
\includecomment{fullproofs}

\newcommand{\cpn}[1]{#1^-}
\newcommand{\cpp}[1]{#1^+}
\newcommand{\mdepth}[1]{\mdepthsymb \left( #1 \right)}

\newcommand{\transa}[1]{\mathtt t_\forall \left( #1 \right)}
\newcommand{\transe}[1]{\mathtt t_\exists \left( #1 \right)}
\newcommand{\transq}[2]{\mathtt t_{#1} \left( #2 \right)}

\newcommand{\translationq}[2]{\mathtt{\mathbf{T}}_{#1}\left( #2 \right)}
\newcommand{\translationmc}[1]{\mathtt{\mathbf{R}}\left( #1 \right)}
\newcommand{\boundvars}[1]{\mathsf{Bound}\left(#1\right)}

\newcommand{\freevars}[1]{\mathsf{Free}\left(#1\right)}

\newcommand{\levelzeroposcopies}[1]{\mathsf P(#1)} 
\newcommand{\levelzeroposcopiesupto}[2]{\mathsf P_{\leq #2}} 
\newcommand{\levelzeronegcopies}[1]{\mathsf N(#1)} 
\newcommand{\levelzeronegcopiesupto}[2]{\mathsf N_{\leq #2}} 
\newcommand{\mdepthsymb}{\mu}
\newcommand{\nblock}[1]{\mathsf{NBlock}\left(#1\right)}
\newcommand{\nbvar}[1]{\mathsf{NQVar}\left(#1\right)}
\newcommand{\boolfun}[2]{\mathbf{f}^{#1}(#2)}
\newcommand{\boolfunname}[1]{\mathbf{f}^{#1}}
\newcommand{\boolfunsemantics}[1]{f^{#1}}
\newcommand{\card}[1]{ {|#1|} }
\newcommand{\infer}[2]      %  \infer{A}{\nec A}  (math mode!)
    {{$\frac{\textstyle #1}{\textstyle #2}$} }

\newcommand{\intFml}[1]{\val \left( #1 \right)}
\newcommand{\lang}[1]{ {\cal L}_{\mathsf{#1}} }
\newcommand{\leqv}{\leftrightarrow}
\newcommand{\limp}{\rightarrow}

\newcommand{\lngth}[1]{\left|\left| #1 \right|\right| }
\renewcommand{\phi}{\varphi}
\newcommand{\propset}     { \mathsf{Var} }
\newcommand{\propsetOf}[1]{ \propset(#1) }

\newcommand{\suchthat}{\ : \ }

\newcommand{\val}{\mathit{V}}
\newcommand{\valb}{\mathit{W}}

\newcommand{\ah}[1]{}
\newtheorem{fact}{Fact}
\newenvironment{pf}{\smallskip\noindent \textsc{Proof.}}
{\hspace*{\fill}\nolinebreak[2]\hspace*{\fill}$\blacksquare$\medskip}

\newcommand{\bit}{\begin{itemize}\item}
\newcommand{\eit}{\end{itemize}}
\newcommand{\ben}{\begin{enumerate}\item}
\newcommand{\een}{\end{enumerate}}

\newcommand{\pspace}{\textsf{\small \textup{PSPACE}}\xspace}
\newcommand{\ptime}{\textsf{\small \textup{P}}}